\documentclass[10pt]{article}
\usepackage{slashed}
\usepackage{amsmath}
\usepackage{amssymb}
\usepackage{axodraw}
\newcommand{\be}{\begin{equation}}
\newcommand{\ee}{\end{equation}}
\newcommand{\ba}{\begin{eqnarray}}
\newcommand{\ea}{\end{eqnarray}}
\newcommand{\rmi}[1]{{\mbox{\scriptsize #1}}}

\newcommand{\tr}{{\rm Tr\,}}
\newcommand{\nn}{\nonumber \\}
\newcommand{\fr}[2]{{\frac{#1}{#2}\,}}
\newcommand{\msbar}{{\overline{\mbox{\rm MS}}}}

\renewcommand{\(}{\left(}
\renewcommand{\)}{\right)}

\newcommand{\e}{\epsilon}

\newcommand{\mLARGE}[1]{\hbox{\LARGE $#1$}}
\def\sumint{\hbox{$\sum$}\!\!\!\!\!\!\!\int}
\newcommand{\I}{{\cal I}^0}
\newcommand{\It}{\widetilde{\cal I}^0}
\newcommand{\Taut}{\widetilde{\mLARGE{\tau}}}
\newcommand{\Mt}{\widetilde{\cal M}}
\newcommand{\N}{{\cal N}}
\renewcommand{\ln}{{\rm ln}}
\newcommand{\mubar}{\bar{\mu}}

\newcommand{\boldmu}{{\mbox{\boldmath$\mu$}}}

\newcommand{\aE}[1]{\alpha_\rmi{E#1}}

\newcommand{\imathb}{i}

\makeatletter
\renewcommand\section{\@startsection {section}{1}{\z@}%
                                   {-5.5ex \@plus -1ex \@minus -.2ex}
                                   {2.3ex \@plus.2ex}%
                                   {\normalfont\large\bfseries}}
\renewcommand\subsection{\@startsection{subsection}{2}{\z@}%
                                     {-3.25ex\@plus -1ex \@minus -.2ex}%
                                     {1.5ex \@plus .2ex}%
                                     {\normalfont\normalsize\bfseries}}
\renewcommand\thesection {\@arabic\c@section}
\renewcommand\thesubsection   {\thesection.\@arabic\c@subsection}
\renewcommand{\@seccntformat}[1]{%
\csname the#1\endcsname.\hspace{1.0em}}
\makeatother

\newcommand{\pic}[1]{\;\parbox[c]{30pt}{\begin{picture}(30,30)(0,0)
\SetWidth{1.0}\SetScale{1.0} #1 \end{picture}}\;}

\newcommand{\picb}[1]{\;\parbox[c]{48pt}{\begin{picture}(45,30)(-9,0)
\SetWidth{1.0}\SetScale{1.0} #1 \end{picture}}\;}
\newcommand{\picc}[1]{\;\parbox[c]{45pt}{\begin{picture}(45,30)(0,0)
\SetWidth{1.0}\SetScale{1.0} #1 \end{picture}}\;}

\def\Lwidth{1}

\def\Agl(#1,#2)(#3,#4,#5){\PhotonArc(#1,#2)(#3,#4,#5){\Lwidth}
{6.283 #3 mul 360 div #4 #5 sub #4 #5 sub mul sqrt mul Ldensity mul}}
\def\Lgl(#1,#2)(#3,#4){\Photon(#1,#2)(#3,#4){\Lwidth}
{#1 #3 sub #1 #3 sub mul #2 #4 sub #2 #4 sub mul add sqrt Ldensity mul}}
\def\Agh(#1,#2)(#3,#4,#5){\DashArrowArc(#1,#2)(#3,#4,#5){1}}
\def\Aagh(#1,#2)(#3,#4,#5){\DashArrowArcn(#1,#2)(#3,#5,#4){1}}
\def\Lgh(#1,#2)(#3,#4){\DashArrowLine(#1,#2)(#3,#4){1}}
\def\Lagh(#1,#2)(#3,#4){\DashArrowLine(#3,#4)(#1,#2){1}}
\def\Ahh(#1,#2)(#3,#4,#5){\DashCArc(#1,#2)(#3,#4,#5){1}}
\def\Lhh(#1,#2)(#3,#4){\DashLine(#1,#2)(#3,#4){1}}
\def\Aqu(#1,#2)(#3,#4,#5){\ArrowArc(#1,#2)(#3,#4,#5)}
\def\Aaqu(#1,#2)(#3,#4,#5){\ArrowArcn(#1,#2)(#3,#5,#4)}
\def\Lqu(#1,#2)(#3,#4){\ArrowLine(#1,#2)(#3,#4)}
\def\Laqu(#1,#2)(#3,#4){\ArrowLine(#3,#4)(#1,#2)}
\def\Aqq(#1,#2)(#3,#4,#5){\CArc(#1,#2)(#3,#4,#5)}
\def\Lqq(#1,#2)(#3,#4){\ArrowLine(#1,#2)(#3,#4)}
\def\Asc(#1,#2)(#3,#4,#5){\ArrowArc(#1,#2)(#3,#4,#5)}
\def\Lsc(#1,#2)(#3,#4){\ArrowLine(#1,#2)(#3,#4)}
\def\DAsc(#1,#2)(#3,#4,#5){\DashCArc(#1,#2)(#3,#4,#5){3}}
\def\DLsc(#1,#2)(#3,#4){\DashLine(#1,#2)(#3,#4){3}}
\def\TAsc(#1,#2)(#3,#4,#5){\SetWidth{2.0}\CArc(#1,#2)(#3,#4,#5)\SetWidth{1.0}}
\def\TLsc(#1,#2)(#3,#4){\SetWidth{2.0}\ArrowLine(#1,#2)(#3,#4)\SetWidth{1.0}}

\makeatletter \@addtoreset{equation}{section} \makeatother
\renewcommand{\theequation}{\arabic{section}.\arabic{equation}}
\makeatletter
\renewcommand\section{\@startsection {section}{1}{\z@}%
                                   {-5.5ex \@plus -1ex \@minus -.2ex}
                                   {2.3ex \@plus.2ex}%
                                   {\normalfont\large\bfseries}}
\renewcommand\subsection{\@startsection{subsection}{2}{\z@}%
                                     {-3.25ex\@plus -1ex \@minus -.2ex}%
                                     {1.5ex \@plus .2ex}%
                                     {\normalfont\normalsize\bfseries}}
\renewcommand\thesection {\@arabic\c@section}
\renewcommand\thesubsection   {\thesection.\@arabic\c@subsection}
\renewcommand{\@seccntformat}[1]{%
\csname the#1\endcsname.\hspace{1.0em}}
\makeatother

\begin{document}


\begin{titlepage}
\begin{flushright}
hep-ph/0305183\\
HIP-2003-30/TH\\
\end{flushright}
\begin{centering}
\vfill

{\large {\bf The pressure of QCD at finite temperatures and chemical potentials}}

\vspace{0.8cm}

{A. Vuorinen\footnote{aleksi.vuorinen@helsinki.fi}}

\vspace{0.8cm}

{\em
Department of Physical Sciences,
Theoretical Physics Division \\
P.O. Box 64,
FIN-00014 University of Helsinki,
Finland\\}

\vspace*{1.4cm}

\end{centering}

\noindent

We compute the perturbative expansion of the pressure of hot QCD to order $g^6\ln\,g$ in the presence of finite quark chemical potentials. In this process we evaluate all two- and three-loop vacuum diagrams of the theory at arbitrary $T$ and $\mu$ and then use these results to analytically verify the outcome of an old order $g^4$ calculation of Freedman and McLerran for the zero-temperature pressure. The results for the pressure and the different quark number susceptibilities at high $T$ are compared with recent lattice simulations showing excellent agreement especially for the chemical potential dependent part of the pressure.

\vfill
\noindent

\vspace*{1cm}

\noindent

\vfill

\end{titlepage}


\section{Introduction}
If either the temperature or the density of strongly interacting matter is increased enough, it undergoes a phase transition from the hadronic phase into deconfined quark-gluon plasma (QGP) at an energy density of approximately 1 GeV/fm$^3$. As one then approaches even higher energies, the value of the gauge coupling keeps decreasing making it possible to start using the machinery of perturbation theory in computing different observables.  The problem of determining the perturbative expansion for the most fundamental thermodynamic quantity, the grand potential $\Omega=-pV$, has been under attack already for more than two decades. It is an especially hot topic today due to the fact that QGP is currently under experimental study in the ongoing heavy-ion experiments at RHIC.

At vanishing chemical potentials, or zero net baryon density, the perturbative series for the pressure has recently been driven to the last fully perturbative order\footnote{At order $g^6$ one runs into infrared problems that can only be solved by non-perturbative methods \cite{linde}.}, $g^6\ln\,g$ \cite{klry,klry2}, following the determination of the contributions of orders $g^2$ \cite{es}, $g^3$ \cite{jk}, $g^4\ln\,g$ \cite{tt}, $g^4$ \cite{az} and $g^5$ \cite{zk,bn1}. At zero temperature and large chemical potentials the expansion is known to $\mathcal{O}(g^4)$ \cite{fmcl}, and at high temperatures but finite chemical potentials to $\mathcal{O}(g^4\ln\,g)$ \cite{tt}. The limit of large chemical potentials and small but non-zero temperatures is at present the least well known; there the only applicable result is of order $g^2$ \cite{es}. In addition to these computations, there have been numerous attempts to determine the pressure using four-dimensional lattice simulations (see e.g. \cite{karsch1,karsch2,fodor,rummu1,gup4}) and the HTL-approximation \cite{birn1,htl2loop,htl2loopq}. In the limit of a large number of flavors the pressure has furthermore recently been non-perturbatively determined both at $\mu=0$ \cite{moore} and $\mu\neq 0$ \cite{ippreb}, and these results have then been used to extract the perturbative expansion of the quantity at large $n_f$ even to order $g^6$ \cite{ippreb}.

The present paper provides a generalization of the order $g^6\ln\,g$ computation \cite{klry} of Kajantie {\em et al.} to finite quark chemical potentials. Using the framework of dimensional reduction and evaluating all one-particle irreducible vacuum diagrams of the theory up to three-loop order, we will derive an analytic expression for the pressure valid at high temperatures and finite chemical potentials. Furthermore, we will determine the different quark number susceptibilities at $\mu=0$, which together with the pressure are compared with recent lattice results of Gavai and Gupta \cite{gup4} showing impressive agreement in particular for the chemical potential dependent part of the pressure. The diagrammatic computations performed here will also be used to tackle the problem of determining the pressure at low temperatures. In particular we will verify the outcome of the well-known $T=0$ computation of Freedman and McLerran \cite{fmcl} and provide a simple analytic value for a poorly-known numerical coefficient appearing in the result.

The paper is organized as follows. In section 2 the general notation is explained, and the necessary special functions are introduced. Section 3 provides then an introduction to dimensional reduction, and the results for the pressure at high $T$ and finite $\mu$ are presented. These results are analyzed in detail in section 4, where we in particular investigate their agreement with lattice data. In section 5 we address the difficult problem of computing the pressure at $T=0$ and show how the result of \cite{fmcl} can be obtained from the computations performed in this
paper. Section 6 is then devoted to addressing the important question of the compatibility of the two results obtained for the pressure at high $T$ and $T=0$, and conclusions are finally drawn in section 7. We leave almost all computational details to the appendices.

\section{Setup and notation}
The theory we consider in this paper is the SU$(N_c)$ Yang-Mills theory coupled to $n_f$ flavors of massless fermions. It is described by the Lagrangian density
\ba
{\cal L}_\rmi{QCD} & = &  \fr14 F_{\mu\nu}^a F_{\mu\nu}^a + \bar\psi\slashed{D}\psi,
\ea
where, as usual,
\ba
F_{\mu\nu}^a &=& \partial_\mu A_\nu^a - \partial_\nu A_\mu^a + g f^{abc} A_\mu^b A_\nu^c,\\
D_\mu &=& \partial_\mu - i g A_\mu \;\;\,=\;\;\, \partial_\mu - i g A_\mu^a T^a,
\ea
and the symbols $T^a$ denote the generators of the fundamental representation of the gauge group. All quark fields have been combined into a multi-component spinor $\psi$, and since flavor is a conserved quantum number of the theory, we may assign an independent chemical potential $\mu_f$ for each $f$. As is customary in finite temperature computations, we will work in Euclidean metric.

In finite temperature field theory the partition function is represented by a functional integral of the exponential of the Euclidean action,
where the usual time integral has been replaced by one over the compact imaginary time $\tau$ ($=x_0$ in Euclidean metric) ranging from $0$ to $1/T$
\ba
{\cal Z}(T,\mu) &=& e^{-\Omega/T}\;\;=\;\; \int \!{\cal D} \phi\, {\rm
exp}\bigg\{\!-\!\int_0^{1/T} \! {\rm d}\tau \!\int \! {\rm d}^{d-1} x\, \big({\cal L}
-\mu {\cal N} \big)\bigg\}.
\label{z}
\ea
The perturbative evaluation of this integral leads to the computation of vacuum Feynman diagrams with Feynman rules closely analogous to the zero-temperature ones. The most important modification is the replacement of the $p_0$ loop integrals by discrete sums over the so-called Matsubara frequencies
\ba
p^\rmi{bos}_0&=&2 n \pi T, \label{mats1}\\
p^\rmi{ferm}_0&=&(2n+1)\pi T-i\mu, \label{mats2}
\ea
where $n$ is an integer. In gauge field theories such as QCD the gauge invariance creates an additional problem, as one needs to restrict the degrees of freedom contributing to the functional integral to the physical ones. In the present paper this is implemented by working in the covariant Feynman gauge throughout the computations.

We end the section by introducing some new notation. The chemical potentials will henceforth usually appear in the dimensionless combinations
\ba
\mubar &\equiv& \mu/(2\pi T), \\
z &\equiv& 1/2-\imathb\mubar,
\ea
and in the context of computing the zero temperature partition function the following abbreviation will be used
\ba
\sum_f\mu_f^2\;\;\,\equiv\;\;\,{\mbox{\boldmath$\mu$}}^2.
\ea
In sums over a single flavor index the subscript $f$ in $\mu_f$ will usually be suppressed.

The momentum integration measure and the notation used for sum-integrals from here onwards are
\ba
\int_p &\equiv& \int\! \fr{{\rm d}^{d} p}{(2\pi)^{d}} \;\;\,=\;\;\, \Lambda^{-2\e}\(\!\fr{e^{\gamma}\bar{\Lambda}^2}{4\pi}\!\!\)^{\!\!\e}\!\!\int\!
\fr{{\rm d}^{d} p}{(2\pi)^{d}}, \\
\sumint_{P/\{P\}} &\equiv& T \sum_{p_0/\{p_0\}} \int_p,
\ea
where $\bar{\Lambda}$ is the $\msbar$ scale, and $p_0$ and $\{p_0\}$ denote the bosonic and fermionic Matsubara frequencies, respectively. We have introduced the unconventional notation $\bar{\Lambda}$ for the $\msbar$ scale in order to avoid confusion with the chemical potentials.

We define some familiar group theory factors by
\ba
C_A \delta^{cd} & \equiv & f^{abc}f^{abd} \;\,\;=\;\,\; N_c \delta^{cd}, \\
C_F \delta_{ij} & \equiv & (T^a T^a)_{ij} \;\,\;=\;\,\; \fr{N_c^2-1}{2N_c}\delta_{ij}, \\
T_F \delta^{ab} & \equiv & \tr T^a T^b \;\,\;=\;\,\; \fr{n_f}{2} \delta^{ab}
\ea
and denote an additional, slightly less well-known one by
\ba
D \delta^{cd} \;\;\,\equiv\;\;\, d^{abc}d^{abd}
\;\;\,=\,\;\; \fr{N_c^2-4}{N_c}\delta^{cd}.
\ea
The dimensions of the adjoint and fermionic representations of the gauge group are naturally
\ba
d_A &\equiv& \delta^{aa} \;\;\,=\;\;\,  N_c^2 - 1, \\
d_F &\equiv& \delta_{ii} \;\;\,=\;\;\, d_A T_F/C_F \;\;\,=\;\;\, N_c n_f.
\ea

For some frequently occurring combinations of special functions we will apply the following abbreviations
\ba
\zeta'(x,y) &\equiv& \partial_x \zeta(x,y), \label{specf1}\\
\aleph(n,w) &\equiv& \zeta'(-n,w)+\(-1\)^{n+1}\zeta'(-n,w^{*}), \\
\aleph(w) &\equiv& \Psi(w)+\Psi(w^*),\label{specf3}
\ea
where $n$ is assumed to be a non-negative integer and $w$ a general complex number. Here $\zeta$ denotes the
Riemann zeta function, and $\Psi$ is the digamma function
\ba
\Psi(w)&\equiv&\fr{\Gamma '(w)}{\Gamma(w)}.
\ea
These functions are analyzed in some detail in appendix D.

\section{The pressure at large $T/\mu$}
\subsection{Dimensional reduction}
In order to compute the partition function of QCD, we need a systematic way of taking into account the contributions of the different momentum scales, as conventional perturbation theory fails already at three-loop order. At high temperatures a physically intuitive solution is offered by dimensional reduction, which is based on the observation that as the temperature is increased, all degrees of freedom except for the ones associated with the zero Matsubara modes of bosonic fields get large effective masses proportional to $T$. They can thus be integrated out leaving us with a three-dimensional effective theory describing the soft scales, where only the bosonic zero modes remain intact. Details of the procedure can be found from \cite{bn1,dr,klrs2}.

The inclusion of the chemical potentials makes the problem of determining the pressure even more complex, as we now need to take into account the effects of these new scales in addition to the usual thermal ones proportional to $T$. Assuming the magnitude of the chemical potentials to be negligible in comparison with $2\pi T$ we may, however, certainly continue using dimensional reduction as a framework \cite{hlp}. This means that the expression for the pressure may be separated into three parts
\ba
p_\rmi{QCD} &=& p_\rmi{E}+p_\rmi{M}+p_\rmi{G},
\ea
corresponding to the contributions of the momentum scales $2\pi T$, $gT$ and $g^2T$, respectively \cite{klry,bn1}. By definition
\ba
p_\rmi{E}\!\(T,\mu\) \;\;\,\equiv\,\;\;
 p_\rmi{QCD}\!\(T,\mu\)-\fr{T}{V} \ln \! \int {\cal D}A_i^a \,
{\cal D}A_0^a \exp\Big\{\!-\!S_\rmi{E}\Big\},
\ea
where $S_\rmi{E}$ is the action of a three-dimensional effective theory with the Lagrangian density \cite{hlp}
\ba
{\cal L}_\rmi{E} \;\;\, = \,\;\; \fr12 \tr F_{ij}^2 + \tr [D_i,A_0]^2 + m_\rmi{E}^2\tr A_0^2 + \fr{\imathb g^3}{3\pi^2}
\sum_f\mu_f \,\tr A_0^3 + \delta{\cal L}_\rmi{E},
\label{leqcd}
\ea
and where the traces are now taken only over the color indices. Similarly, $p_\rmi{M}$ and $p_\rmi{G}$ are defined by
\ba
p_\rmi{M}\!\(T,\mu\) &\equiv& p_\rmi{QCD}\!\(T,\mu\)-p_\rmi{E}\!\(T,\mu\)-\fr{T}{V} \ln  \!\int {\cal D}A_i^a
\exp\Big\{\!-\!S_\rmi{M}\Big\} \nn
&\equiv& p_\rmi{QCD}\!\(T,\mu\)-p_\rmi{E}\!\(T,\mu\) -p_\rmi{G}\!\(T\), \\
{\cal L}_\rmi{M} & = & \fr12 \tr F_{ij}^2 + \delta{\cal L}_\rmi{M}.
\ea
The gauge coupling constants of the two effective theories, $g_\rmi{E}$ and $g_\rmi{M}$, appear in the covariant derivatives above, and operators contributing to the partition function starting at $\mathcal{O}(g^6)$ or higher have been assembled to the terms $\delta{\cal L}_\rmi{E}$ and $\delta{\cal L}_\rmi{M}$. The question, at which values of the chemical potentials we may trust results obtained using dimensional reduction, is examined quantitatively in \cite{hlp} and is also briefly discussed in the section 6 of this paper.

At leading order the different parts contribute to the pressure as $p_\rmi{E}\sim g^0$, $p_\rmi{M}\sim g^3$ and
$p_\rmi{G}\sim g^6\ln\,g$. The first of these functions can be obtained by computing the strict perturbation expansion of
the pressure in the full theory, i.e. by evaluating all 1PI vacuum diagrams of four-dimensional QCD without applying
any form of resummation. The two other ones are then available by evaluating the perturbative expansions of the partition
functions of the effective theories \cite{klry2}, the parameters of which must, however, be determined through the full theory \cite{hlp,klrs}. Following the notation of \cite{klry} and using results from \cite{klry,bir}, these statements can be summarized by writing
\ba
\label{pe}
 \fr{p_\rmi{E}(T,\mu)}{T\Lambda^{-2\e}} & = & T^3 \bigg[\aE{1}
 + g^2
 \Big(\aE{2} + {\cal O}(\epsilon)\Big)
 + \frac{g^4}{(4\pi)^2}
 \Big(\aE{3} + {\cal O}(\epsilon)\Big)+ {\mathcal O}(g^6)
 \bigg], \\
 m_\rmi{E}^2 & = & T^2 \bigg[ g^2
 \Big( \aE{4} +
 \aE{5} \epsilon + {\cal O}(\epsilon^2) \Big)
 + \frac{g^4}{(4\pi)^2}
 \Big( \aE{6} +
 {\cal O}(\epsilon) \Big)+ {\mathcal O}(g^6)  \bigg], \hspace*{0.5cm}  \\
 g_\rmi{E}^2 & = & T \bigg[ g^2 + \frac{g^4}{(4\pi)^2}
 \Big( \aE{7} + {\cal O}(\epsilon) \Big)+ {\mathcal O}(g^6) \bigg], \\
 \frac{p_\rmi{M}(T,\mu)}{T \Lambda^{-2 \epsilon} } & = &
 \frac{1}{(4\pi)}
 d_A  m_\rmi{E}^3
 \bigg[\fr13 + {\cal O}(\epsilon) \bigg] \nn
 & + &
 \frac{1 
 }{(4\pi)^2}
 d_A C_A
 g_\rmi{E}^2 m_\rmi{E}^2
 \bigg[-\frac{1}{4\epsilon} - \fr34 -\ln\frac{\bar{\Lambda}}{2 m_\rmi{E}}
 + {\cal O}(\epsilon) \bigg] \nn
 & + &
 \frac{1 
 }{(4\pi)^3}
 d_A C_A^2
 g_\rmi{E}^4 m_\rmi{E}
 \bigg[-\frac{89}{24} - \fr16 \pi^2 + \frac{11}{6}\ln\,2
 + {\cal O}(\epsilon) \bigg] \nn
 & + &
 \frac{1 
 }{(4\pi)^4}
 d_A C_A^3
 g_\rmi{E}^6\, \ln\frac{\bar{\Lambda}}{2 m_\rmi{E}}
 \bigg[ \fr{43}{4}-\fr{491}{768}\pi^2
  + {\cal O}(\epsilon)
 \bigg]    \nn
& + &
\frac{1 
}{(4\pi)^4}
d_A D T_F^2
g_\rmi{E}^6\,\ln\frac{\bar{\Lambda}}{2 m_\rmi{E}}
\bigg[ -\fr{16}{3n_f^2}\bigg(\sum_f \mubar\bigg)^{\!2}
+ {\cal O}(\epsilon) \bigg]+ {\mathcal O}(g^6), \label{pm}
\ea
\ba
m_\rmi{M} &=& C_A g_\rmi{M}^2+ {\mathcal O}(g^3),    \\
g_\rmi{M}^2 &=& g_\rmi{E}^2+ {\mathcal O}(g^3), \\
 \frac{p_\rmi{G}(T)}{T \Lambda^{-2 \epsilon} } &=&
 d_A C_A^3 \frac{g_\rmi{M}^6}{(4\pi)^4}\, \ln\frac{\bar{\Lambda}}{2m_\rmi{M}}
 \Big[\fr{43}{12}-\fr{157}{768}\pi^2 + {\mathcal O}(\epsilon) \Big]+ {\mathcal O}(g^6), \label{pg}
\ea
where $g$ is the renormalized gauge coupling of quantum chromodynamics and the matching coefficients $\alpha$ are left to be determined. Apart from modifying the values of these coefficients, the effects of finite $\mu$ can only be seen in the
appearance of the last term in Eq. (\ref{pm}). In particular, one should notice that $p_\rmi{G}$ depends only on $T$ at the order considered here.


\def\Elmeri(#1,#2,#3){{\pic{#1(15,15)(15,0,180)%
 #2(15,15)(15,180,360)%
 #3(0,15)(30,15)}}}

\def\Petteri(#1,#2,#3,#4,#5,#6){\pic{#3(15,15)(15,-30,90)%
 #1(15,15)(15,90,210)%
 #2(15,15)(15,210,330) #5(2,7.5)(15,15) #6(15,15)(15,30) #4(15,15)(28,7.5)}}

\def\Jalmari(#1,#2,#3,#4,#5,#6){\picc{#1(15,15)(15,90,270)%
 #2(30,15)(15,-90,90) #4(30,30)(15,30) #3(15,0)(30,0) #5(15,0)(15,30)%
 #6(30,30)(30,0) }}

\def\Oskari(#1,#2,#3,#4,#5,#6,#7,#8){\picc{#1(15,15)(15,90,270)%
 #2(30,15)(15,-90,90) #4(30,30)(15,30) #3(15,0)(30,0) #6(15,0)(15,15)%
 #5(15,15)(15,30) #8(30,30)(30,15) #7(30,15)(30,0) }}

\def\Sakari(#1,#2,#3){\picb{#1(15,15)(15,30,150)%
#1(15,15)(15,210,330) #2(0,15)(7.5,-90,90) #2(0,15)(7.5,90,270) %
#3(30,15)(7.5,-90,90) #3(30,15)(7.5,90,270) }}

\def\Maisteri(#1,#2){\picb{#1(15,15)(15,0,150)%
#1(15,15)(15,210,360) #2(0,15)(7.5,-90,90) #2(0,15)(7.5,90,270) #1(37.5,15)(7.5,0,360) }}

\begin{figure}[t]
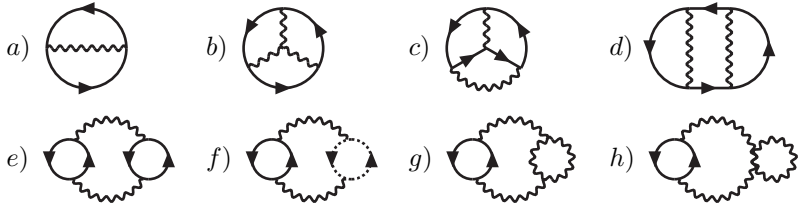

\centering
\ba \nonumber
\begin{array}{llll}
a)~ \Elmeri(\Asc,\Asc,\Lgl) & b)~
\Petteri(\Asc,\Asc,\Asc,\Lgl,\Lgl,\Lgl)
& c)~
\Petteri(\Asc,\Agl,\Asc,\Lsc,\Lsc,\Lgl)
& d)~
\Jalmari(\Asc,\Asc,\Lsc,\Lsc,\Lgl,\Lgl)
\nn
\nn
e)~
\Sakari(\Agl,\Asc,\Asc)
& f)~
\Sakari(\Agl,\Asc,\Agh)
& g)~
\Sakari(\Agl,\Asc,\Agl)
& h)~
\Maisteri(\Agl,\Asc)
\nn
\end{array}
\ea
\caption[a]{The two- and three-loop fermionic diagrams of the full theory contributing to the values of $\aE{2}$ and $\aE{3}$. The solid, wiggly and dotted lines stand respectively for the quark, gluon and ghost propagators.}
\end{figure}

With the exception of $\aE{3}$ and $\aE{5}$, the results for the matching coefficients can be immediately extracted from
\cite{klry,es,hlp,bir}. To get $\aE{5}$ we furthermore merely need to evaluate the one-loop gluon polarization tensor
to ${\mathcal O}(\e)$ in the limit of vanishing external momenta, which is a simple computation. The calculation of
$\aE{3}$ is, on the other hand, already a considerably more laborious task, as it involves computing all
three-loop vacuum diagrams of the theory. They have so far only been evaluated at vanishing chemical potentials \cite{az} and temperatures \cite{fmcl}, and the generalization of these calculations to finite $\mu$ and $T$ is the topic of appendices A and B. The relevant fermionic
two- and three-loop diagrams are depicted in Fig. 1, and the results for the matching coefficients can be found from below.

In order to write the perturbation theory result for the pressure in the familiar form of a power series in the coupling
constant, we simply need to add together Eqs. (\ref{pe}), (\ref{pm}) and (\ref{pg}) and expand the result in $g$. Up to
order $g^6\ln\,g$ the outcome reads
\ba
 \frac{p_\rmi{QCD}(T,\mu)}{T^4 \Lambda^{-2 \epsilon} } & = &
 \frac{p_\rmi{E}(T,\mu) + p_\rmi{M}(T,\mu) + p_\rmi{G}(T)}{T^4 \Lambda^{-2 \epsilon} } \nn
 & = &
 g^0 \bigg\{ \aE{1} \bigg\}
  +
 g^2 \bigg\{ \aE{2} \bigg\}
  +
 \frac{g^3}{(4\pi)}
 \bigg\{ \frac{d_A}{3} \aE{4}^{3/2} \bigg\} \nn
 & + &
 \frac{g^4}{(4\pi)^2} \bigg\{
 \aE{3} - d_A C_A
 \bigg[
 \aE{4} \bigg(
 \frac{1}{4\epsilon} + \fr34 + \ln\frac{\bar{\Lambda}}{2 g T \aE{4}^{1/2}}
 \bigg)
 + \fr14 \aE{5} \bigg] \bigg\} \nn
 & + &  \frac{g^5}{(4\pi)^3} \bigg\{ d_A \aE{4}^{1/2} \bigg[
 \fr12 \aE{6} - C_A^2
 \bigg(
 \frac{89}{24} + \frac{\pi^2}{6} - \frac{11}{6} \ln\,2
 \bigg)
 \bigg] \bigg\}  \nn\label{pres}
 & + &  \frac{g^6}{(4\pi)^4} \bigg\{d_AC_A\Big(\aE{6}+\aE{4}\aE{7}\Big)\ln\Big[g \aE{4}^{1/2}\Big]
+\fr{16}{3n_f^2}\bigg(\sum_f \mubar\bigg)^{\!2}\,d_A D T_F^2 \,\ln\Big[g\aE{4}^{1/2}\Big]\\
&-&
8\,d_A C_A^3\bigg[
 \(\fr{43}{32}-\fr{491}{6144}\pi^2\)\ln\Big[g \aE{4}^{1/2}\Big]
 + \(\fr{43}{48}-\fr{157}{3072}\pi^2\) \ln\Big[g C_A^{1/2}\Big] \bigg]\bigg\}+\mathcal{O}(g^6),
\nonumber
\ea
where the pole of $\aE{3}$ exactly cancels the $1/\e$ term appearing in the order $g^4$ contribution. As there is an unknown $\mathcal{O}(g^6)$ contribution missing from the result, there is an ambiguity in choosing the coefficients inside the logarithms of the order $g^6\ln\,g$ terms\footnote{Varying the coefficients inside the logarithms amounts to varying the magnitude of the undetermined order $g^6$ term.}. The current choice is, however, well-motivated from the effective theory point of view; the logarithms now appear in the same form they emerged from the three-dimensional Feynman diagrams.

\subsection{The matching coefficients}
Given in terms of the special functions and group theory factors defined in the previous section, the results for the matching coefficients $\alpha$ read
\ba
\aE{1} &=& \fr{\pi^2}{45}\fr{1}{n_f}\!\sum_f\bigg\{d_A+\bigg(\fr{7}{4} + 30\mubar^2 + 60\mubar^4\bigg)d_F\bigg\},
\label{ae1} \\
\aE{2} &=& -\fr{d_A}{144}\fr{1}{n_f}\!\sum_f\bigg\{C_A + \fr{T_F}{2}\(1+12\mubar^2\)\(5+12\mubar^2\)\!\bigg\}, \\
\aE{3} &=& \fr{d_A}{144}\Bigg[\fr{1}{n_f}\!\sum_f\bigg\{C_A^2\bigg(\fr{12}{\e}
+\fr{194}{3}\ln\fr{\bar{\Lambda}}{4\pi T} + \fr{116}{5} + 4\gamma -\fr{38}{3}\fr{\zeta'(-3)}{\zeta(-3)} +
\fr{220}{3}\fr{\zeta'(-1)}{\zeta(-1)}\bigg) \nn
&+& C_A T_F\bigg( \!12\(1+12\mubar^2\)\fr{1}{\e} +
\bigg(\fr{169}{3}+600\mubar^2-528\mubar^4\bigg)\ln\fr{\bar{\Lambda}}{4\pi T} +\fr{1121}{60} + 8\gamma \nn
&+& 2\(127+48\gamma\)\mubar^2 - 644\mubar^4
+ \fr{268}{15}\fr{\zeta '(-3)}{\zeta(-3)} + \fr{4}{3}\(11+156\mubar^2\)\fr{\zeta'(-1)}{\zeta(-1)} \nn
&+& 24\Big[52\,\aleph(3,z)
+ 144\imathb\mubar\,\aleph(2,z)+\(17-92\mubar^2\)\aleph(1,z)+4\imathb\mubar\,\aleph(0,z)\Big]\bigg) \nn
&+& C_F T_F \bigg(\fr{3}{4}\(1+4\mubar^2\)\(35+332\mubar^2\)-24\(1-4\mubar^2\)\fr{\zeta'(-1)}{\zeta(-1)} \nn
&-& 144\Big[12\imathb\mubar\,\aleph(2,z)-2\(1+8\mubar^2\)\aleph(1,z)
-\imathb\mubar\(1+4\mubar^2\)\aleph(0,z)\Big]\bigg) \nn
&+& T_F^2 \bigg(\fr{4}{3}\(1+12\mubar^2\)\(5 + 12\mubar^2\)\ln\fr{\bar{\Lambda}}{4\pi T}
+ \fr{1}{3}+4\gamma + 8\(7+12\gamma\)\mubar^2 + 112\mubar^4
- \fr{64}{15}\fr{\zeta'(-3)}{\zeta(-3)} \nn
&-& \fr{32}{3}\(1+12\mubar^2\)\fr{\zeta'(-1)}{\zeta(-1)}
- 96\Big[8\,\aleph(3,z) + 12\imathb\mubar\,\aleph(2,z) - 2\(1+2\mubar^2\)\aleph(1,z)
- \imathb\mubar\,\aleph(0,z)\Big]\bigg)\bigg\} \nn
&+& 288\,T_F^2\fr{1}{n_f^2}\sum_{f\,g}\bigg\{2\(1+\gamma\)\mubar_f^2\mubar_g^2
-\Big[\aleph(3,z_f+z_g)+\aleph(3,z_f+z_g^*) \nn
&+&4\imathb\mubar_f\Big(\aleph(2,z_f+z_g) + \aleph(2,z_f+z_g^*)\Big)
- 4\mubar_g^2\,\aleph(1,z_f) -\(\mubar_f+\mubar_g\)^2\aleph(1,z_f+z_g) \nn
&-&\(\mubar_f-\mubar_g\)^2\aleph(1,z_f+z_g^*)
-4\imathb\mubar_f\mubar_g^2\,\aleph(0,z_f) \Big]\bigg\}\Bigg], \label{alphae3} \\
\aE{4} &=& \fr{1}{3}\fr{1}{n_f}\!\sum_f\Big\{C_A+T_F\(1+12\mubar^2\)\Big\}, \\
\aE{5} &=& \fr{1}{3}\fr{1}{n_f}\!\sum_{f}
\bigg\{2\,C_A\bigg(\ln\fr{\bar{\Lambda}}{4\pi T} + \fr{\zeta'(-1)}{\zeta(-1)}\bigg)\nn
&+& T_F \bigg(\(1+12\mubar^2\)\(2\,\ln\fr{\bar{\Lambda}}{4\pi T}+1\) + 24\,\aleph(1,z)\bigg)\bigg\},
\ea
\ba
\aE{6} &=& \fr{1}{9}\fr{1}{n_f}\!\sum_f\bigg\{C_A^2\bigg(22\,\ln\fr{e^{\gamma}\bar{\Lambda}}{4\pi T}+5\bigg)
+C_AT_F\bigg(2\(7+132\mubar^2\)\ln\fr{e^{\gamma}\bar{\Lambda}}{4\pi T}+9+132\mubar^2+8\gamma + 4\,\aleph(z)\bigg) \nn
&-& 18\,C_F T_F\Big(1+12\mubar^2\Big) - 4\,T_F^2\Big(1+12\mubar^2\Big)\bigg(2\,\ln\fr{\bar{\Lambda}}{4\pi T} - 1 - \aleph(z) \bigg)\bigg\}, \\
\aE{7} &=& \fr{1}{3}\fr{1}{n_f}\sum_{f}\bigg\{C_A\bigg(22\,\ln\fr{e^{\gamma}\bar{\Lambda}}{4\pi T}
+1\bigg) - 4\,T_F\(2\,\ln\fr{\bar{\Lambda}}{4\pi T}-\aleph(z)\)\bigg\}.
\ea
Combined with Eq. (\ref{pres}), this is the main result of the paper.

\section{Lattice tests}
As there unfortunately is no experimental data available for the pressure in the QGP phase, the results derived for this quantity in the
previous section can only be compared to other analytic computations\footnote{A recent large-$n_f$ computation by Ipp and
Rebhan provides an accurate numerical check for some of the results of this paper. For details, see section 3 of
\cite{ippreb}.} or to lattice simulations. In particular, it is
very interesting to investigate, to what accuracy we can reproduce the results of the various lattice studies that have
been performed for the pressure and the different quark number susceptibilities. In this section these comparisons will
be made, and it will furthermore be studied, how rapidly the perturbative series for the different quantities converge.

\subsection{The pressure}
It has been observed that at vanishing chemical potentials the order $g^6\ln\,g$ perturbative result for the pressure is
well compatible with four-dimensional lattice simulations, but that the eventual determination of the yet unknown $g^6$
term may still change the situation dramatically \cite{klry}. The perturbative result varies largely from order to order
and even at $\mathcal{O}(g^4)$ its behavior as a function of temperature still bears no resemblance to the lattice
predictions, even though the order $g^2$ result gives a relatively good estimate for the quantity. For the $\mu$
dependent part of the pressure we will see a significant improvement in these convergence properties.

\begin{figure}[t]

\centerline{\epsfxsize=7.3cm\epsfysize=6.3cm \epsfbox{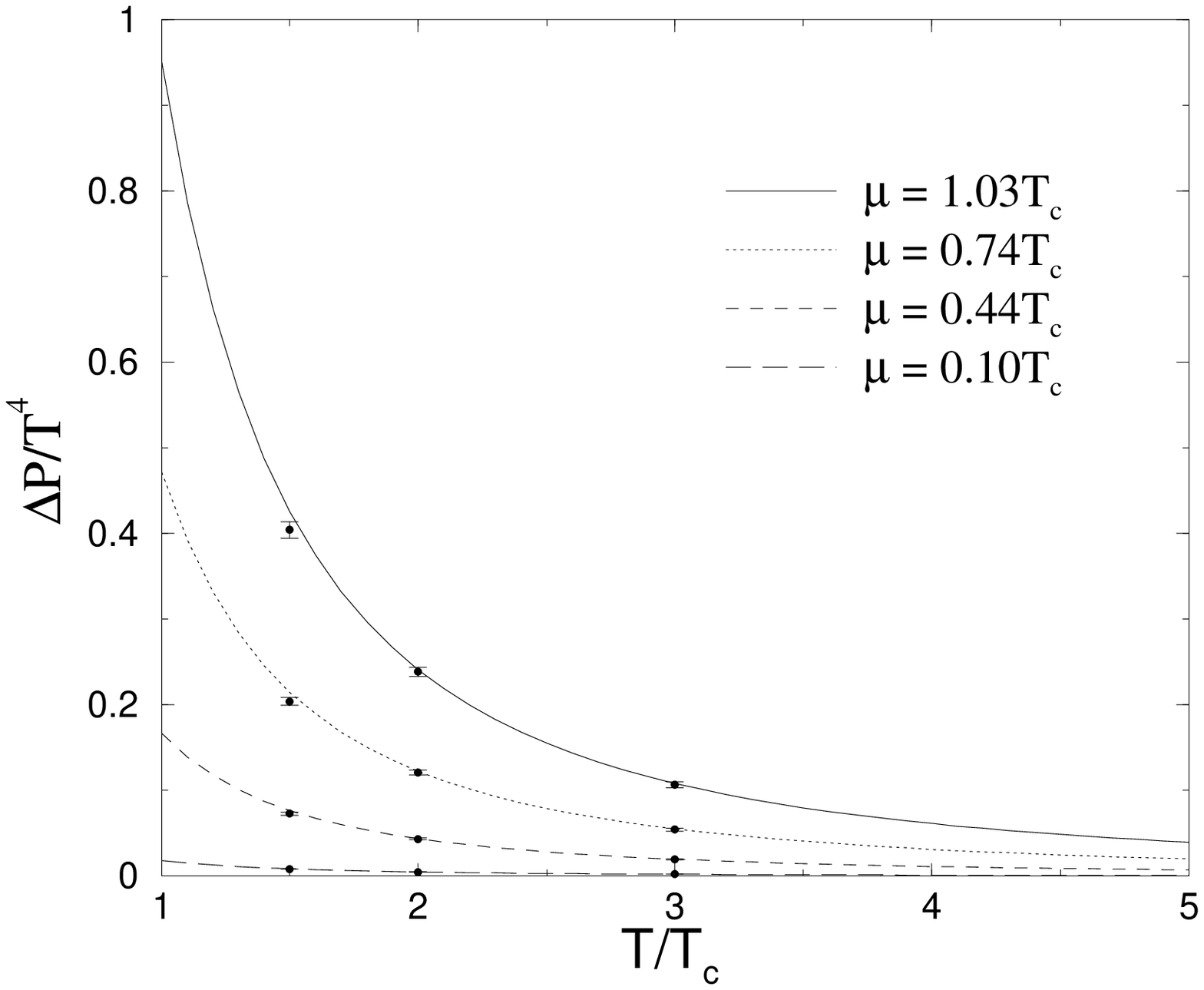}\;\;\;\;\;\;\epsfxsize=7.3cm\epsfysize=6.3cm \epsfbox{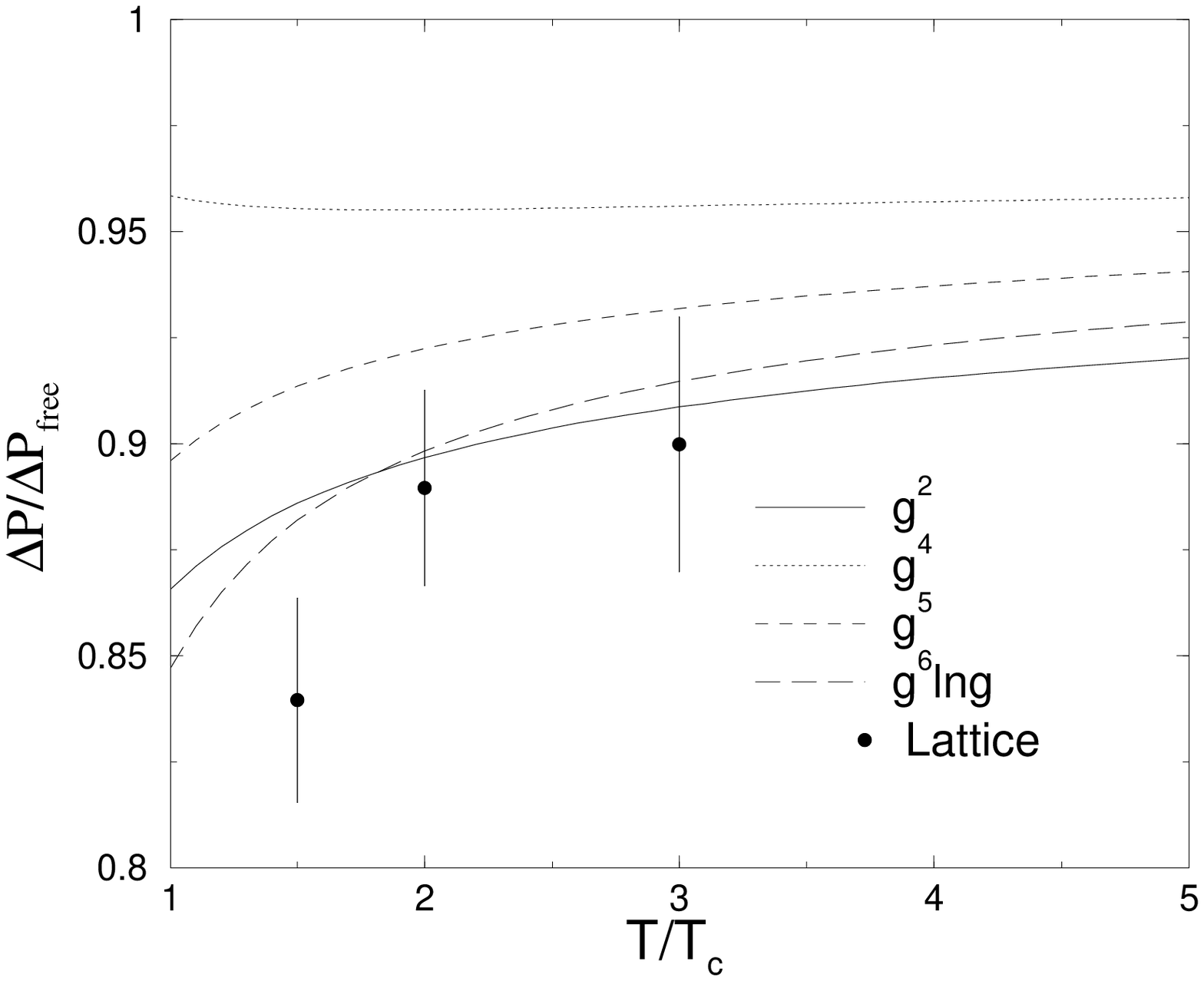}}

\caption[a]{The perturbative and lattice \cite{gup4} results for $\Delta P$ plotted as functions of $T/T_c$. On the left
$\mu$ has been given different values, and on the right the different perturbative orders for the $\mu=0.44T_c$ case are
shown normalized to the free theory result. The result $T_c/\Lambda_{{\overline{\rm MS}}}|_{n_f=0}=1.15$ \cite{gup1} and the $\mu=0$
convention $\bar{\Lambda}=6.742T$ \cite{klrs} have been applied here.}
\end{figure}

Extracting the quantity
\ba
\Delta P(T,\mu)&=&p_\rmi{QCD}(T,\mu)-p_\rmi{QCD}(T,0) \label{deltap}
\ea
from Eq. (\ref{pres}), the $\mu$ dependence of the pressure can be directly compared with recent lattice studies \cite{gup4}, where $\Delta P$ has been computed in quenched QCD assuming two light flavors of quarks, $u$ and $d$, at equal chemical potentials. In Fig. 2 (left) this lattice data is plotted alongside with the perturbative result, which has been obtained by setting all explicit factors of $n_f$ to zero in order to match the quenched approximation. One observes that already at temperatures $T\approx 2T_c$ the perturbative results lie well within the error bars of the lattice datapoints and that the differences between subsequent perturbative orders are very small.

As we can see from Fig. 2 (right), the picture is qualitatively similar to the $\mu=0$ case in the sense that the leading correction to the free theory result already gives a good estimate for the quantity in question. The next perturbative orders then
make the situation worse until at $\mathcal{O}(g^5)$ one again starts approaching the lattice results. The main
difference between the two cases is simply that $\Delta P$ is a much more strongly perturbative quantity: for it even the
free theory result falls within 10$\%$ of the lattice data and the relative magnitudes of the perturbative corrections
are considerably smaller than for the $\mu=0$ pressure.

\begin{figure}[t]

\centerline{\epsfxsize=7.3cm\epsfysize=6.3cm \epsfbox{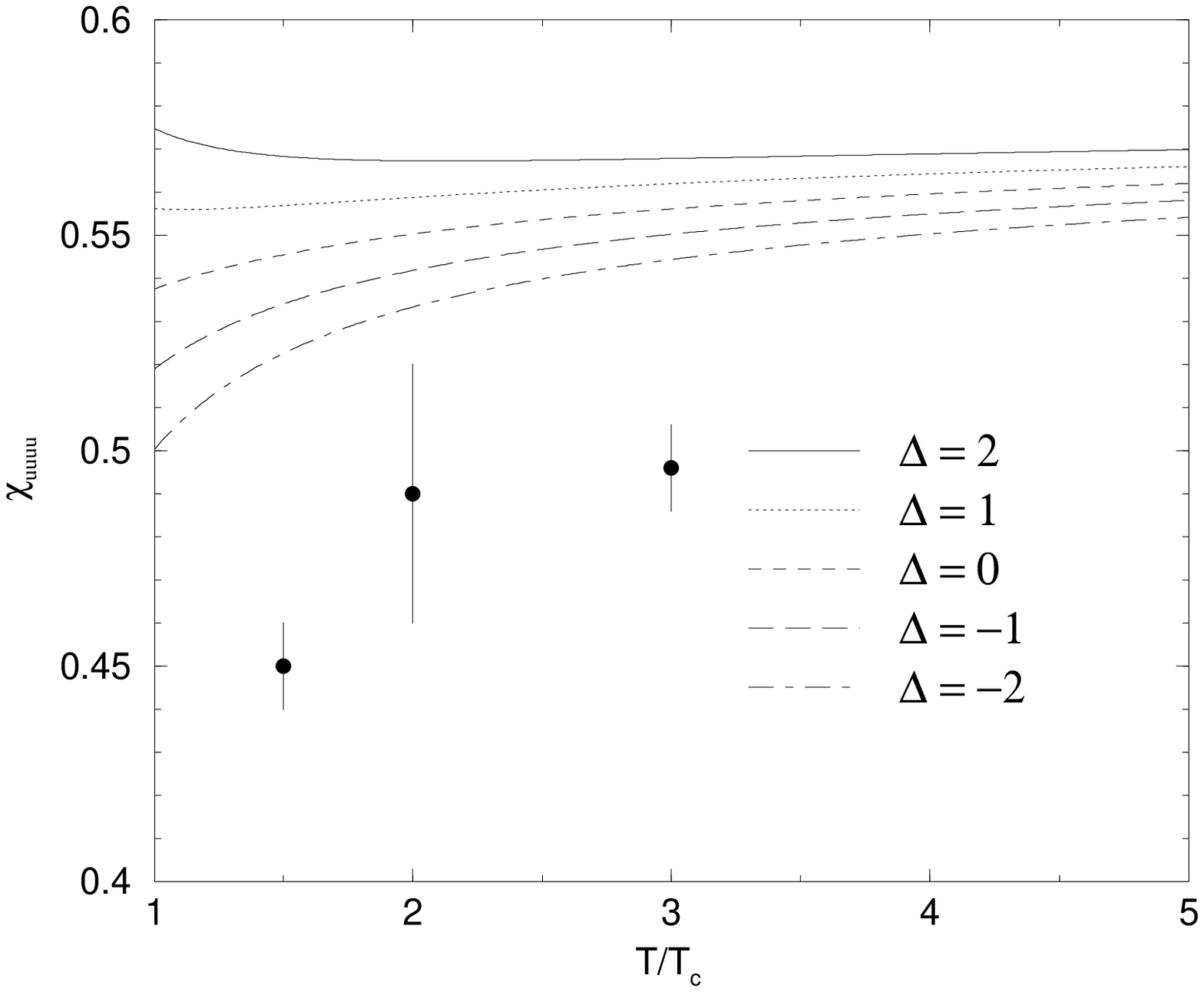}\;\;\;\;\;\;\epsfxsize=7.3cm\epsfysize=6.3cm \epsfbox{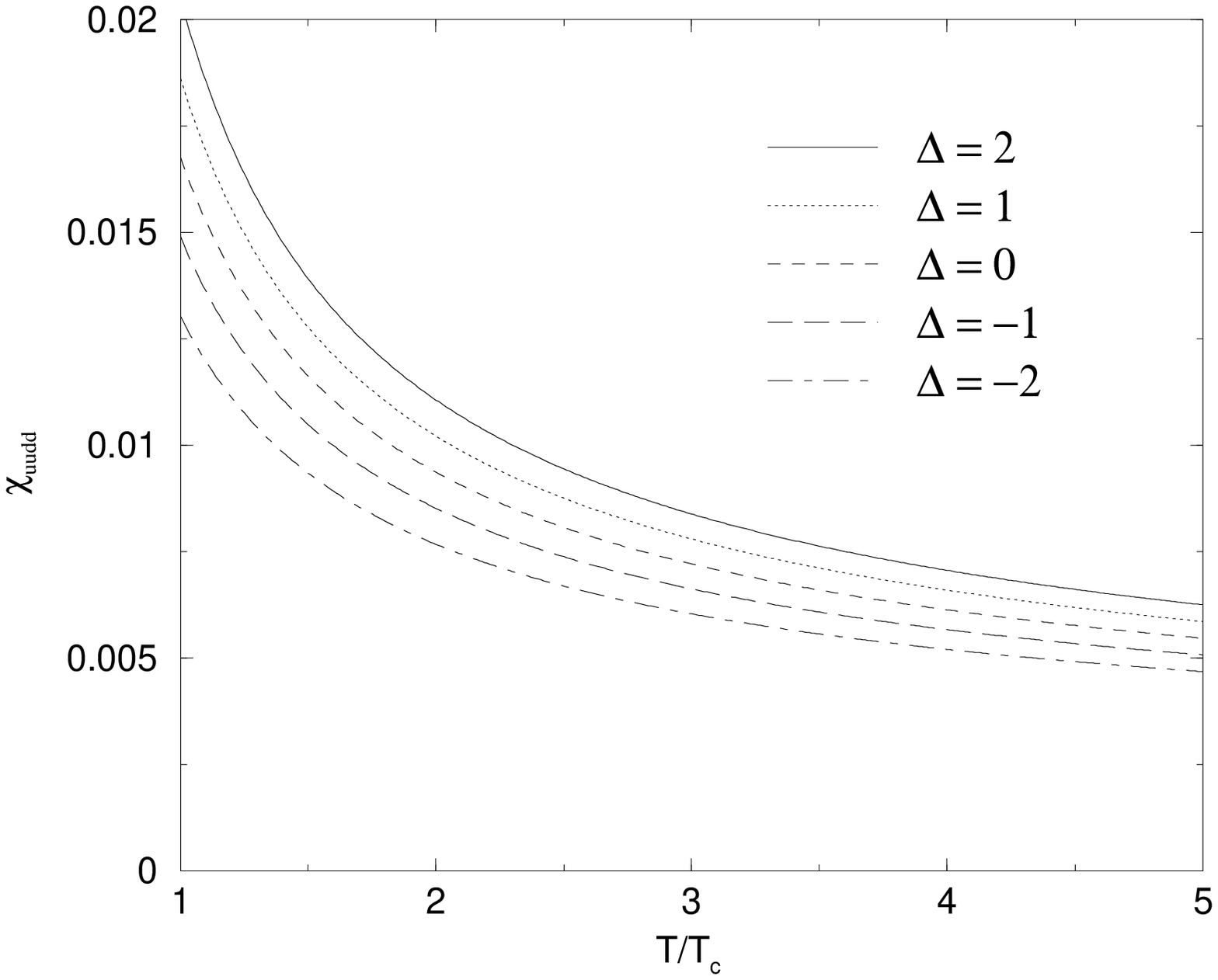}}

\caption[a]{Perturbative results for $\chi_{uuuu}$ and $\chi_{uudd}$ at $n_f=0$ plotted with lattice results from
\cite{gup4}. The curves corresponding to different values of $\Delta$ show the expected effect of the yet undetermined
$g^6$ term of the perturbative expansion (for details, see \cite{av}). Again $T_c/\Lambda_{{\overline{\rm MS}}}|_{n_f=0}=1.15$
\cite{gup1} and $\bar{\Lambda}=6.742T$ \cite{klrs}.}
\end{figure}

\subsection{Quark number susceptibilities}
Apart from analyzing $\Delta P$ directly, there are other, more effective ways to investigate the chemical potential
dependence of the results. To make full use of the large amount of lattice data existing at $\mu=0$ (see e.g.
\cite{gup4,gup2,gup3,aleij}) we may use Eq. (\ref{pres}) and the results of appendix D to compute the different quark
number susceptibilities
\ba
\chi_{ijk...}&\equiv&\fr{\partial^n p}{\partial\mu_i \partial\mu_j \partial\mu_k ...}
\ea
in this limit. The linear (i.e. second order) susceptibilities have already been considered both in the framework of
ordinary perturbation theory \cite{av} and in the HTL approximation \cite{bir,mustafa,birmus} with the result that only
the diagonal ones are accurately predicted by perturbation theory. For the linear non-diagonal susceptibility the
perturbative results were found to be several orders of magnitude larger than the lattice ones \cite{bir}. One should,
however, note that for the non-diagonal susceptibilities there is considerable disagreement between the different
lattice approaches (see the discussion in \cite{aleij}).

A similar behavior can be observed when studying the nonlinear quark number susceptibilities. In Fig. 3 we have plotted
the susceptibilities $\chi_{uuuu}$ and $\chi_{uudd}$ at $n_f=0$\footnote{In this context the $n_f=0$ limit of the
perturbative result is understood to be taken only after the necessary differentiations of the pressure with respect to
$\mu_f$ have been carried out. The pressure at $n_f=0$ is defined in an analogous manner.}, and comparing with the
quenched QCD results of \cite{gup4} we again see that the diagonal quantity is satisfactorily produced by perturbation
theory but that the prediction for the non-diagonal one is too large by more than a factor of 1000. This apparent
disagreement is, however, not unexpected, as even the different lattice results for the non-diagonal susceptibilities
differ from each other. Furthermore, the perturbative expansions for the nonlinear susceptibilities start only at
relatively high orders ($\mathcal{O}(g^6\ln\,g)$ for $\chi_{ud}$, $\mathcal{O}(g^3)$ for $\chi_{uudd}$), and it is
therefore entirely possible that large cancellations will occur as one drives perturbation theory even further. The
situation is completely different in the case of the the diagonal susceptibilities, for which the free theory result
already gives the correct order of magnitude of the results. E.g. for $\chi_{uuuu}$ one obtains from Eq. (\ref{ae1})
\ba
\chi_{uuuu}&\equiv&\fr{\partial^4 p}{\partial\mu_u^4}\;\;=\;\;\fr{6}{\pi^2}+\mathcal{O}(g^2)\;\;\simeq\;\;
0.61+\mathcal{O}(g^2),
\ea
which is in good agreement with the lattice data (see Fig. 3.a).

For $\Delta P$, and hence for the susceptibilities, the order $g^6$ term in the perturbative series contains no
non-perturbative contributions and is therefore in principle straightforwardly obtainable, unlike the corresponding term
in the expansion of the $\mu=0$ pressure \cite{klry}. This computation may already be enough to improve the
perturbative predictions for the non-diagonal susceptibilities significantly, but will have practically no effect on the
already good convergence properties of the chemical potential dependent part of the pressure. As has been pointed out in
\cite{gup4}, the effects of the nonlinear susceptibilities on $\Delta P$ are negligible for small values of the chemical
potentials, and the quantity is almost solely determined by the linear diagonal susceptibilities.

\section{The pressure at $T=0$}
The zero-temperature pressure of QCD was first computed to $\mathcal{O}(g^4)$ a long time ago \cite{fmcl} in a lengthy
calculation involving numerical integrations. Using the analytic results for three-loop diagrams derived in the present
paper this result can be straightforwardly analytically reproduced. The computation is divided into two distinct parts:
the results for the graphs that remain infrared convergent at $T=0$ may be immediately continued to this limit, but in
addition an infinite set of IR divergent ring diagrams must be summed over explicitly. Analogously to the use of the
three-dimensional effective theories in section 3, this resummation is necessary to ensure that the contributions of
all momentum scales are properly accounted for. Only the results of the computation are given below, while the details
are left to appendix E.

\subsection{IR convergent diagrams}
At $T=0$ QCD pressure gets contributions only from the fermionic graphs, i.e. from the diagrams of Fig. 1. Aside from
the IR divergent graph $I_e$ we obtain using the results of appendices A - D
\ba
p_\rmi{1}&\equiv& \(\aE{1} + I_a +  I_b + I_c + I_d
+ I_f + I_g + I_h\){\mbox{\large{$\mid$}}}_{T=0}\nonumber
\ea
\ba
&=&\fr{1}{4\pi^2}\sum_f\mu^4\bigg\{\fr{N_c}{3}-d_A\bigg(\!\fr{g(\bar{\Lambda})}{4\pi}\!\!\bigg)^{\!\!2}
-d_A\bigg[\fr{2n_f}{3\e}+\fr{2}{3}\(11N_c+4n_f\)\ln\fr{\bar{\Lambda}}{\mu}\nn
&+&\fr{17}{4}\fr{1}{N_c} + \fr{1}{36}\(415-264\,\ln\,2\)N_c+\fr{2}{3}\(5-4\,\ln\,2\)n_f\bigg]
\bigg(\!\fr{g}{4\pi}\!\!\!\bigg)^{\!\!4}\bigg\},
\label{pt0a}
\ea
where the renormalization of the gauge coupling has been taken into account.

\begin{figure}[t]
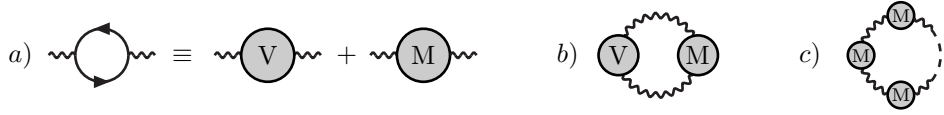

\centering
\ba \nonumber
\begin{array}{lll}
a)~ \pic{\Lgl(0,15)(10,15)%
 \Asc(20,15)(10,0,180) \Asc(20,15)(10,180,360) \Lgl(30,15)(40,15)}\;\;\;\;\equiv\; \pic{\Lgl(0,15)(10,15)%
\Lgl(30,15)(40,15)\GCirc(20,15){10}{0.8} \Text(20,15)[c]{V}} \;\;\;\;+
\pic{\Lgl(0,15)(10,15)\GCirc(20,15){10}{0.8}\Text(20,15)[c]{M}%
 \Lgl(30,15)(40,15)} \;\;\;\;\;\;\;\;\;\;
&b)~ \picb{\Agl(15,15)(15,30,150)%
\Agl(15,15)(15,210,330) \GCirc(0,15){7.5}{0.8}\Text(0,15)[c]{V} %
\GCirc(30,15){7.5}{0.8}\Text(30,15)[c]{M}}
\;\;\;\;\;\;
&c)~ \picb{\Agl(15,15)(15,110,160)%
\Agl(15,15)(15,200,250) %
\GCirc(0,15){5}{0.8} \GCirc(15,30){5}{0.8} \GCirc(15,0){5}{0.8} \Agl(15,15)(15,40,70)\Agl(15,15)(15,290,320)
\DAsc(15,15)(15,-40,40)\Text(15,0)[c]{{${\mbox{\scriptsize{M}}}$}}
\Text(15,30)[c]{{${\mbox{\scriptsize{M}}}$}}\Text(0,15)[c]{{${\mbox{\scriptsize{M}}}$}} }
\nn
\end{array}
\ea
\caption[a]{a) The fermionic part of the one-loop gluon polarization tensor divided into its vacuum and matter parts. \\
b) The diagram $I_e'$ contributing to the zero temperature pressure. \\
c) The generic form of the ring diagrams contributing to $p_\rmi{3}$.}
\end{figure}

\subsection{Ring diagrams}
In order to obtain the correct expression for the zero-temperature pressure up to order $g^4$ we need to add to Eq. (\ref{pt0a})
the contributions of all ring diagrams of the type $I_e$. Individually these graphs are infrared divergent but when summed
together they give a finite contribution to the pressure starting at $\mathcal{O}(g^4\ln\,g)$. Separating the fermionic
part of the one-loop
gluon polarization tensor into its vacuum ($T=\mu=0$) and matter (vacuum subtracted) parts as in Fig. 4.a. we observe
that to order $g^4$ only the diagram $I_e'$ of Fig. 4.b. and the ring sum of Fig. 4.c. need to be computed. The reason for this is that starting at four-loop order the diagrams with at least one vacuum insertion only contribute at $\mathcal{O}(g^6\ln\,g)$ or higher, and the corresponding three-loop diagram with two vacuum insertions naturally vanishes at $T=0$.

A straightforward computation performed in appendix E shows that the diagram $I_e'$ gives the following contribution to the pressure
\ba
p_\rmi{2}\;\;\,=\;\;\,\fr{d_A n_f}{4\pi^2}\sum_f\mu^4\bigg\{\fr{2}{3\e}
+4\,\ln\fr{\bar{\Lambda}}{\mu}+\fr{52}{9}-4\,\ln\,2\bigg\}\bigg(\!\fr{g}{4\pi}\!\!\!\bigg)^{\!\!4}. \label{pqcd2res}
\ea
The summation of the ring diagrams was, on the other hand, first performed in \cite{fmcl} and is reproduced in appendix E
following in most parts the treatment of the original paper. The result of this computation reads
\ba
p_\rmi{3}
&=&-\fr{d_A}{4\pi^2}({\mbox{\boldmath$\mu$}}^2)^2\bigg(\!\fr{g}{4\pi}\!\!\!\bigg)^{\!\!4}\bigg\{4\,\ln\,\fr{g}{4\pi} -
\fr{22}{3} +\fr{16}{3}\,\ln\,2\(1-\ln\,2\) + \delta+ \fr{2\pi^2}{3} \nn
&+&\fr{16}{3}\ln\,2\sum_f\fr{\mu_f^4}{({\mbox{\boldmath$\mu$}}^2)^2}+\fr{F({\mbox{\boldmath$\mu$}})}{({\mbox{\boldmath$\mu$}}^2)^2}\bigg\},
\label{pt03}
\ea
where we have defined
\ba
F({\mbox{\boldmath$\mu$}})&=& -2{\mbox{\boldmath$\mu$}}^2\sum_f \mu^2\,\ln\,\fr{\mu^2}{{\mbox{\boldmath$\mu$}}^2}
+\fr{2}{3}\sum_{f>g}\bigg\{(\mu_f-\mu_g)^2\ln\,\fr{|\mu_f^2-\mu_g^2|}{\mu_f\mu_g} \nn
&+& 4\mu_f\mu_g(\mu_f^2+\mu_g^2)\ln\,\fr{(\mu_f+\mu_g)^2}{\mu_f\mu_g}-(\mu_f^4-\mu_g^4)\ln\,\fr{\mu_f}{\mu_g}\bigg\}.
\ea
The constant $\delta$ possesses the integral representation
\ba
\delta&\equiv&\fr{16}{\pi}\int_0^{\pi/2}{\rm d}x\sin^2x\Bigg\{\fr{(1-x\cot\,x)^2}{\sin^4x}\ln\,\fr{1-x\cot\,x}{\sin^2x}
+ \fr{1}{2}\(1-\fr{1-x\cot\,x}{\sin^2x}\)^2\ln\,\bigg[1-\fr{1-x\cot\,x}{\sin^2x}\bigg]\Bigg\} \nn
&\simeq&-0.85638320932694280684831023291594035884727909711135760899309086726768550829, \label{deltadef}
\ea
which we have not been able to evaluate in closed form. Even all attempts of expressing its numerical value in terms of
the most common natural constants using the PSLQ algorithm \cite{pslq} have been unsuccessful. Finding the correct
basis of constants for $\delta$ seems to be a very non-trivial task.

\subsection{The result at $T=0$}
Eqs. (\ref{pt0a}), (\ref{pqcd2res}) and (\ref{pt03}) together verify the well-known result of \cite{fmcl}
\ba
\label{pt0res}
p_\rmi{QCD}(T=0)&=&p_1+p_2+p_3+\mathcal{O}(g^6\ln\,g)\nn
&=&\fr{1}{4\pi^2}\Bigg(\sum_f \mu^4\bigg\{\fr{N_c}{3}-d_A\bigg(\!\fr{g}{4\pi}\!\!\!\bigg)^{\!\!2}
-d_A\bigg(\!\fr{g}{4\pi}\!\!\!\bigg)^{\!\!4}\bigg[\fr{2}{3}\(11N_c-2n_f\)\ln\fr{\bar{\Lambda}}{\mu} +
\fr{16}{3}\ln\,2 \nn
&+&\fr{17}{4}\fr{1}{N_c} + \fr{1}{36}\(415-264\,\ln\,2\)N_c-\fr{4}{3}\(\fr{11}{6}-\ln\,2\)n_f\bigg]\bigg\} \\
&-& d_A\bigg(\!\fr{g}{4\pi}\!\!\!\bigg)^{\!\!4}\bigg\{\Big(4\,\ln\,\fr{g}{4\pi} -
\fr{22}{3} +\fr{16}{3}\,\ln\,2\(1-\ln\,2\) + \delta+ \fr{2\pi^2}{3}\Big)({\mbox{\boldmath$\mu$}}^2)^2+F({\mbox{\boldmath$\mu$}})\bigg\}\Bigg)+\mathcal{O}(g^6\ln\,g). \nonumber
\ea
In particular, we have here obtained an analytic value $\fr{17}{4}$ for the coefficient of the $\fr{1}{N_c}$ term, which
was previously known only numerically with considerable error bars. When comparing Eq. (\ref{pt0res}) with the result of
\cite{fmcl}, one should notice that there the authors work in the momentum subtraction scheme, in which the gauge coupling constant is related
to the one of the $\msbar$ scheme through the equation
\ba
\fr{g_{{\mbox{\rm \tiny{MOM}}}}^2}{4\pi^2}\;\;\,=\;\;\,\fr{g_{{\overline{\mbox{\rm \tiny{MS}}}}}^2}{4\pi^2}\bigg\{1+\(\fr{151}{144}N_c-\fr{5}{18}n_f\)\fr{g_{\overline{\mbox{\rm \tiny{MS}}}}^2}{4\pi^2}\bigg\}.
\ea

\section{Compatibility of the results at large $\mu/T$}

\begin{figure}[t]
\centerline{\epsfxsize=7.3cm\epsfysize=6.2cm \epsfbox{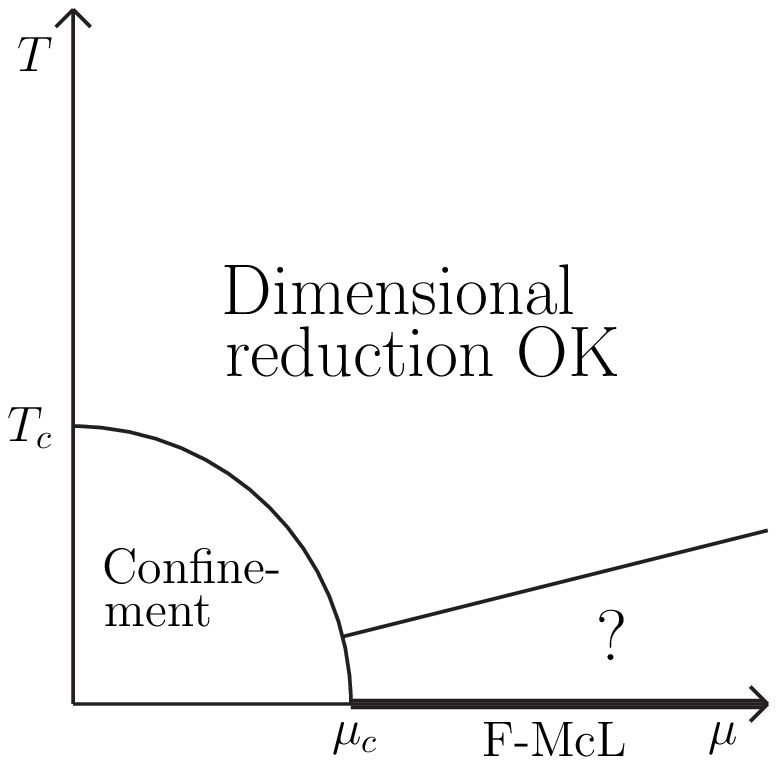}\;\;\;\;\;\;\epsfxsize=7.3cm \epsfbox{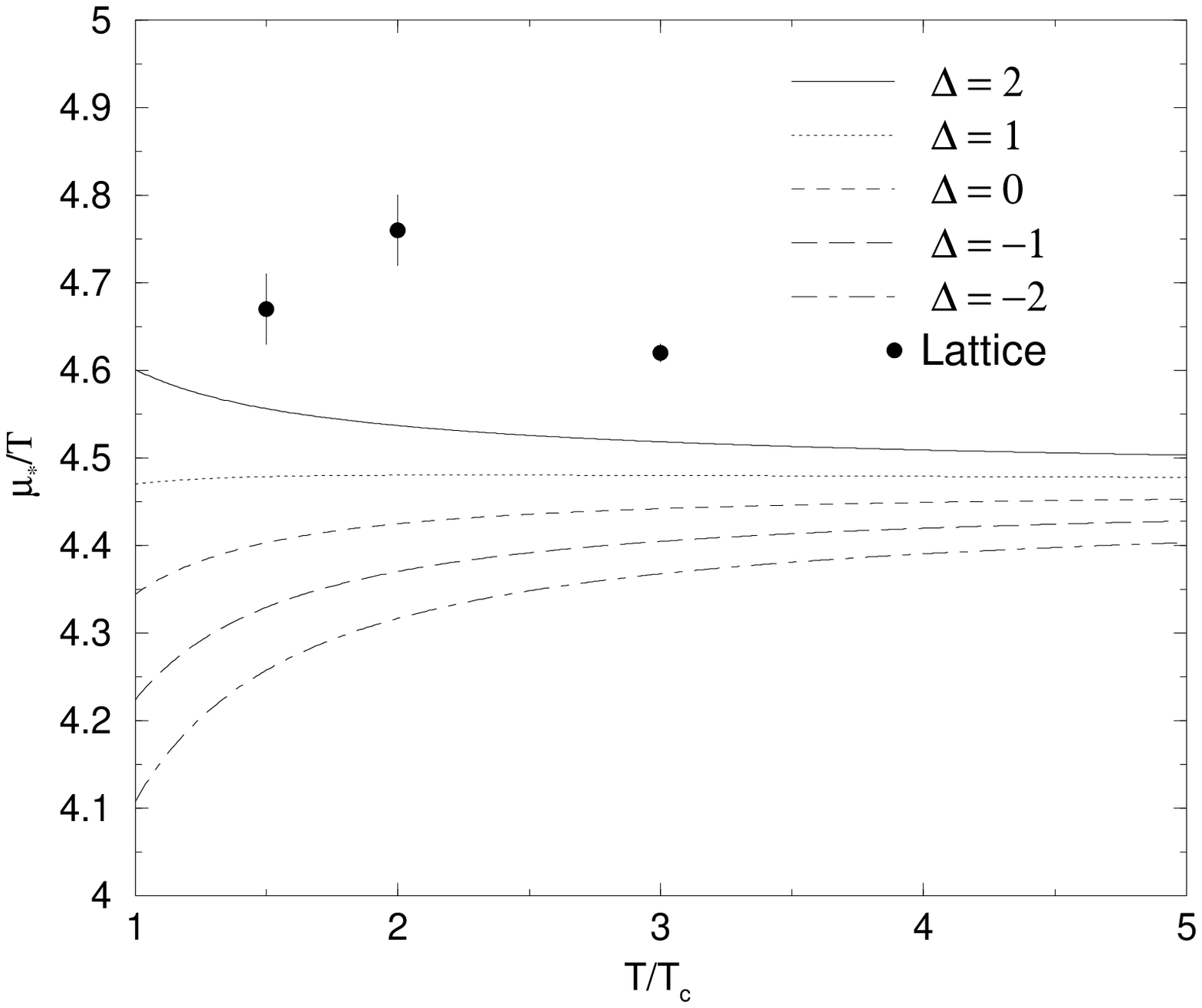}}
\caption[a]{On the left a sketch of the $\mu$-$T$ plane, where F-McL stands for the $T=0$ result, Eq. (\ref{pt0res}).
Next to it the perturbative result for the quantity $\mu_*$ plotted for $n_f=0$ together with lattice data from \cite{gup4}. Once again $T_c/\Lambda_{{\overline{\rm MS}}}|_{n_f=0}=1.15$ \cite{gup1} and $\bar{\Lambda}=6.742T$ \cite{klrs}.}
\end{figure}

In the sections 3 and 5 of the present paper we have derived perturbative results for the QCD pressure in the two limits of
high temperature and small chemical potentials and $T=0$ and large $\mu$. Determining the exact region of applicability
for the first one is a nontrivial task, as it has not been studied analytically, how the appearance of the additional
scales
$\mu_f$ affects the validity of dimensional reduction. In \cite{hlp} it has, however, been estimated based on numerical
results for correlation lengths that the method applies as long as $\mu_f\lesssim 4T$ for all flavors. This
seems physically very
reasonable, since one certainly expects the framework of dimensional reduction to be unaltered, if the values of the
chemical potentials are much smaller that the thermal scale $2\pi T$. Perhaps not surprisingly, roughly the same result
was obtained in \cite{gup4} for the quantity
\ba
\fr{\mu_*}{T} &\equiv&\sqrt{\fr{12\chi_{uu}}{T^2\chi_{uuuu}}}
\ea
describing the highest value of chemical potentials, for which the linear susceptibilities accurately produce the
$\mu$-dependent part of the pressure, $\Delta P$. In Fig. 5 (right) the perturbation theory and lattice results for $\mu_*$ are
plotted as functions of temperature showing reasonable agreement. The overall scale of the results is again given by the
free theory expression, which now reads
\ba
\fr{\mu_*}{T} &=& \sqrt{2}\pi + \mathcal{O}(g^2)\;\;\simeq\;\;4.4 + \mathcal{O}(g^2).
\ea

There remains a region on the $\mu$-$T$ plane between the lines $T=0$ and $T=\mu /4$, where the
perturbative expansion of the pressure is only available to $\mathcal{O}(g^2)$. To obtain an order $g^4$ result valid
throughout the deconfined phase, one would have to perform an explicit summation of all the bosonic and fermionic ring
diagrams at an arbitrary temperature, as adding a mass term for the zero mode of $A_0$ to the free
Lagrangian of the theory would not lead to the expected result in the limit $T\rightarrow 0$. Even though this procedure
is enough to produce the correct $\mathcal{O}(g^4)$ result for the pressure at high $T$ \cite{az}, it does not work at
low temperatures due to the nontrivial structure of the gluon polarization tensor at a vanishing temperature and external
momentum (see appendix E). The separation of the zeroth Matsubara mode of the $A_0$ field is furthermore clearly
inconsistent, if $\mu \gg \pi T$.

The interesting limit of small but non-zero temperatures can in any case be formally taken also in Eq. ($\ref{pres}$),
even though it is already beforehand understood that an unphysical logarithmic divergence of the type
$\ln(T/\bar{\Lambda})$ will appear there. As $T$ approaches zero, it is natural to
investigate, how the difference of Eqs. (\ref{pres}) and (\ref{pt0res}) behaves as a function of the
chemical potentials, as it gives the
magnitude of the terms that have been neglected in deriving Eq. (\ref{pres}) but are necessary to obtain the correct
$T=0$ pressure. This quantity has been plotted in Fig. 6 (right) for different temperatures alongside with the
corresponding curves for the pressure. In this figure we assume two flavors of massless quarks at equal chemical
potentials, and the scale parameter has somewhat arbitrarily been chosen to be
\ba
\bar{\Lambda}&=&2\pi\sqrt{T^2+\fr{\mu^2}{2\pi^2}}
\ea
in analogy with the free theory pressure, Eq. (\ref{ae1}).

\begin{figure}[t]
\centerline{\epsfxsize=7.3cm \epsfbox{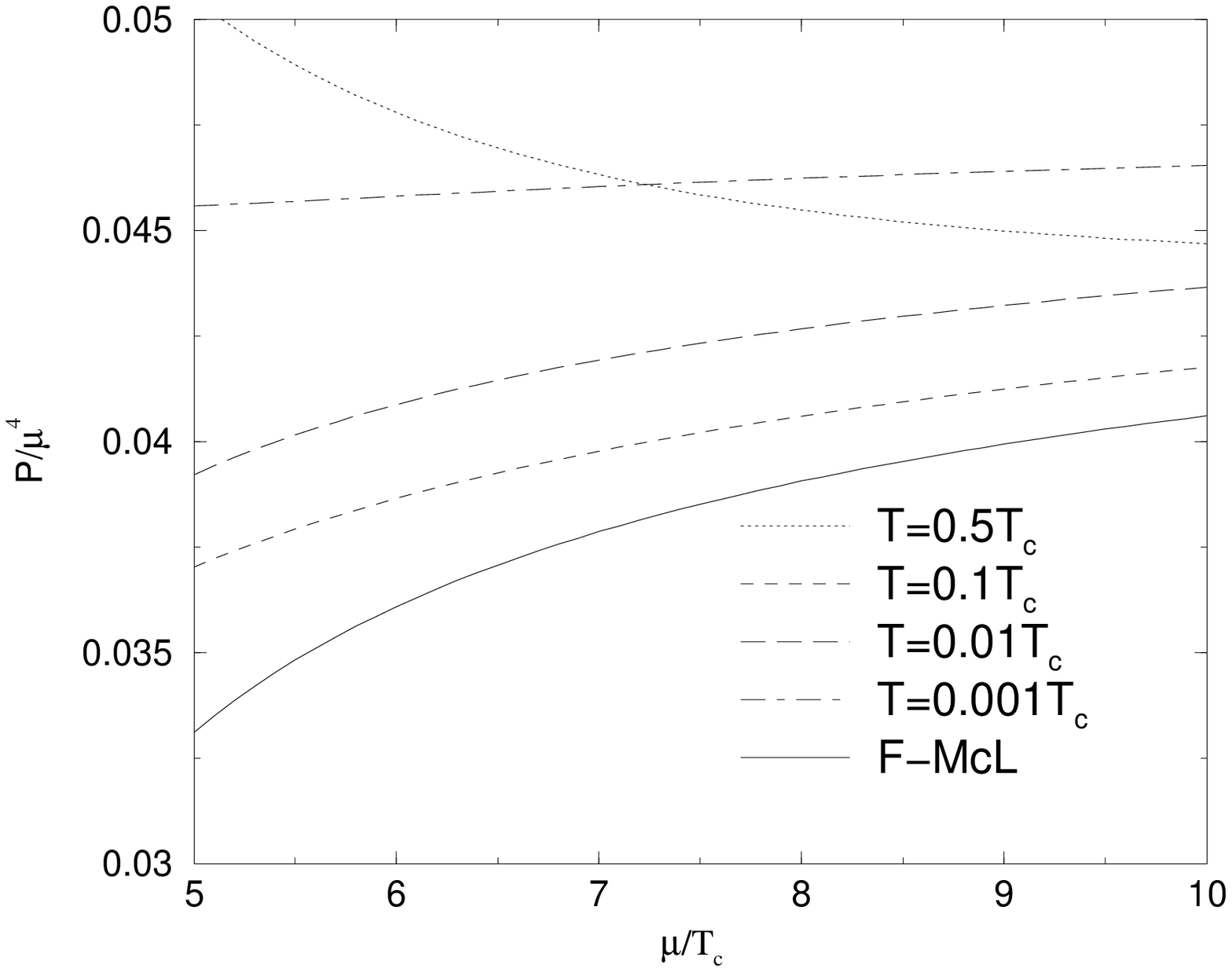}\;\;\;\;\;\;\epsfxsize=7.3cm \epsfbox{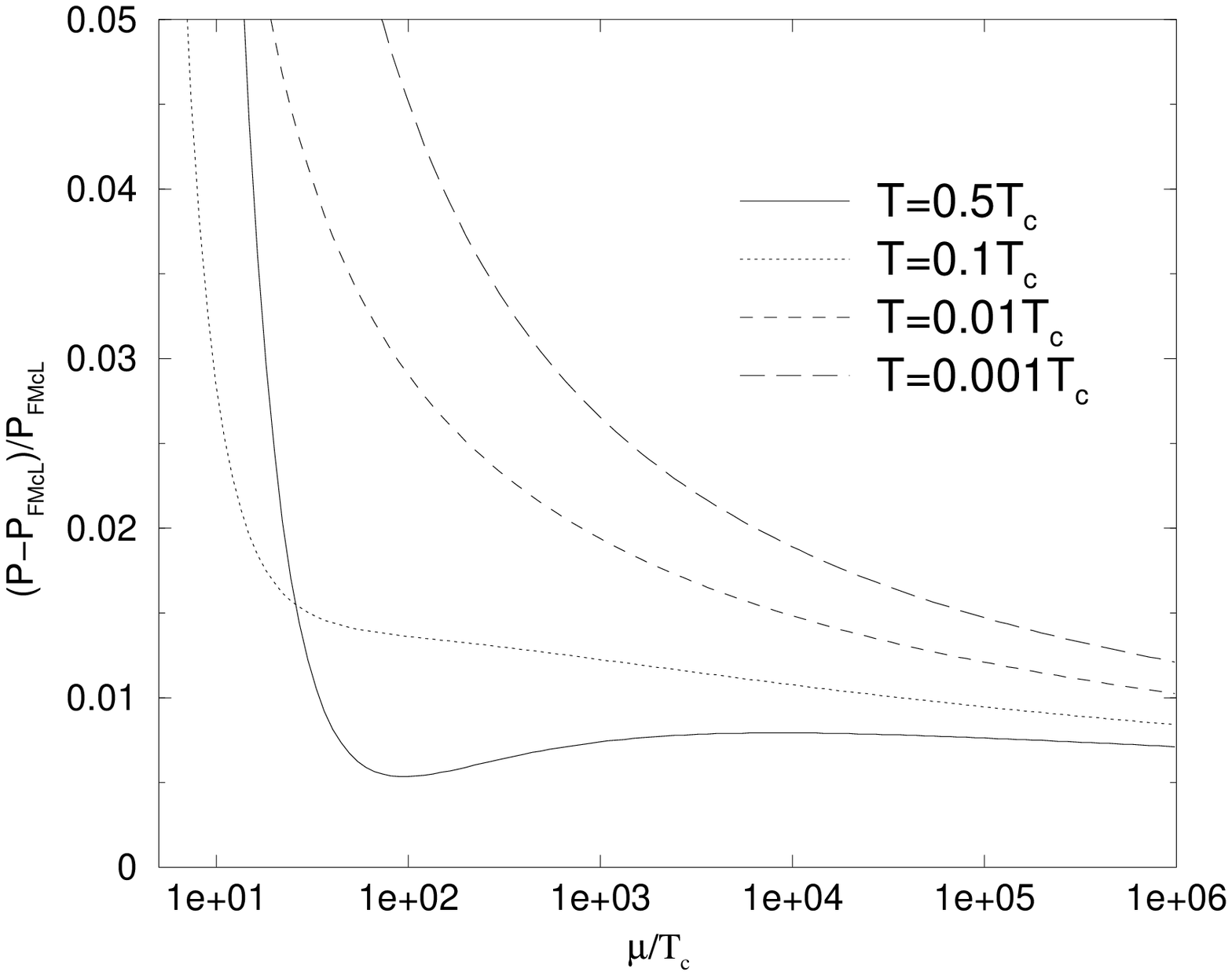}}
\caption[a]{On the left the behavior of Eq. (\ref{pres}) as a function of the chemical potentials is portrayed for different temperatures alongside with the $T=0$ result, Eq. (\ref{pt0res}). On the right the difference of these results is shown on a logarithmic scale appropriately normalized. The result $T_c/\Lambda_{{\overline{\rm MS}}}|_{n_f=2}=0.49$ \cite{gup1} has been applied here.}
\end{figure}

The result shown in Fig. 6 is remarkable. It seems that as we approach the zero-temperature limit, the curves
corresponding to Eq. (\ref{pres}) smoothly approach the one describing Eq. (\ref{pt0res}), until only at very
low temperatures $T\ll T_c$ the logarithmic divergences start increasing the gap. This observation suggests that the
magnitudes of the terms not present in Eq. (\ref{pres}) are small and one is able to use this result throughout
the deconfined phase with the exception of a narrow strip near the $T=0$ line. We in particular notice that a large
value of $\mu/T$ does not itself appear to spoil the applicability of Eq. (\ref{pres}) and that the restriction
$\mu\lesssim 4T$ may therefore perhaps be loosened.

Despite all the optimism, we must be very careful in interpreting Fig. 6: only one special configuration of
chemical potentials has been analyzed so far, and the reasoning presented above is merely of qualitative nature. It is furthermore clear that dimensional reduction cannot be reliably applied in the limit of small temperatures. The good compatibility of Eqs. (\ref{pres}) and (\ref{pt0res}) is most likely simply a consequence of the fact that in both of them the numerically dominant part comes from the strict perturbation expansion of the pressure, Eqs. (\ref{ae1}) - (\ref{alphae3}). Another aspect to keep in mind is that perturbative computations such as the one presented in this paper can never produce the rich phase structure of the `condensed matter QCD' \cite{rajwil} found in this region of the $\mu$-$T$ plane. Thus the applicability of the present results is in any case very limited there.

\section{Conclusions}
In this paper we have improved the perturbative expansion of the pressure of hot and dense QCD by
three orders, as the $g^4$, $g^5$ and $g^6\ln\,g$ terms in the series have been determined. The
crucial step in the computation was the analytic evaluation of all three-loop 1PI vacuum diagrams of the theory at arbitrary $T$ and $\mu$,
which were also used to derive an order $g^4$ result for the zero-temperature pressure. Finally, it was argued based on
a qualitative analysis of our results that the perturbative expansion of the pressure is now converging relatively well
on almost the whole $\mu$-$T$ plane.

There is, however, a large amount of work left to be done. At high temperatures and small chemical potentials
one clearly needs to determine the next $\mathcal{O}(g^6)$ term in the perturbative expansion, as this order contains the
first non-perturbative contributions to the pressure and furthermore has a potentially very significant impact on the
behavior of the result. Another challenge can be found in improving the present embarrassing record in the limit of large
chemical potentials and small but non-zero temperatures; a generalization of the order $g^4$ result at $T=0$ to non-zero
temperatures would certainly be welcome, even if its numerical effect on the present results turned out to be small.

\section*{Acknowledgements}
I would like to thank Keijo Kajantie for suggesting the topic and for all his help and advice throughout the project. In
addition I want to express my gratitude to Antti Gynther, Mikko Laine, Tuomas Lappi, Anton Rebhan, Kari Rummukainen, York Schr\"oder
and Mikko Veps\"al\"ainen for their valuable comments and suggestions. A special thanks belongs to Sourendu Gupta for the
many enlightening discussions we had on the topic. This work has been supported by the Magnus Ehrnrooth Foundation, the
V\"ais\"al\"a Foundation and the Academy of Finland, Contract no. 77744.

\newpage

\appendix

\renewcommand{\thesection}{Appendix~\Alph{section}}
\renewcommand{\thesubsection}{\Alph{section}.\arabic{subsection}}
\renewcommand{\theequation}{\Alph{section}.\arabic{equation}}

\newcommand{\logT}{\ln\frac{\bmu}{4 \pi T}}
\newcommand{\logz}[1]{\frac{\zeta'(-#1)}{\zeta(-#1)}}

\section{Vacuum diagrams}
In order to complete the calculations of section 3, we still need to evaluate the two- and three-loop 1PI vacuum diagrams of
QCD. Since the purely bosonic graphs are unaffected by the finiteness of the chemical potentials and have been
computed already previously \cite{az}, one may restrict the treatment here to the fermionic diagrams of Fig. 1. The
calculations will be performed keeping both the temperature and the chemical potentials arbitrary, which together with the relations listed in appendix D enables one to immediately continue the results to the limits of small $T$ and small $\mu$.

Let us start by defining a set of `master' sum-integrals
\ba
\label{ints1}
{\cal I}_{n}^m &\equiv& \sumint_P \fr{\(p_0\)^m}{\(P^2\)^n}, \\
\widetilde{\cal I}_{n}^m &\equiv& \sumint_{\{P\}} \fr{\(p_0\)^m}{\(P^2\)^n}, \\
\widetilde{\mLARGE{\tau}} &\equiv& \sumint_{\{PQ\}}
\fr{1}{P^2 Q^2 \(P-Q\)^2}, \\
\widetilde{\mLARGE{\tau}}' &\equiv& \sumint_{\{PQ\}}
\fr{p_0}{P^2 Q^2 \(P-Q\)^4}, \\
\widetilde{\cal M}_{m,n} &\equiv& \sumint_{\{PQR\}}
\fr{1}{P^2Q^2\(R^2\)^{m}\big(\(P-Q\)^2\big)^n\(P-R\)^2\(Q-R\)^2}, \\
{\cal N}_{m,n} &\equiv& \sumint_{\{PQ\}R}
\fr{1}{P^2Q^2\(R^2\)^{m}\big(\(P-Q\)^2\big)^n\(P-R\)^2\(Q-R\)^2},
\label{ints2}
\ea
where the notation is adopted from \cite{bn1}. Using the finite temperature Feynman rules (see e.g. \cite{leb}) in the Feynman gauge and taking
advantage of the fact that the purely bosonic version of $\widetilde{\mLARGE{\tau}}$ vanishes at $\mathcal{O}(\e^0)$
\cite{az}, we may write the diagrams of Fig. 1 in terms of the sum-integrals of Eqs. (\ref{ints1}) - (\ref{ints2}). This
computation is lengthy and requires the application of numerous tricks such as linear changes of integration momenta but is nevertheless of
straightforward nature and is therefore not reproduced here. The result of the procedure, correct to $\mathcal{O}(\e^0)$,
reads
\ba
\label{diaga}
I_a &=& -\(1-\e\)d_A g^2 \sum_f \Big\{\It_1\(\It_1-2\I_1\)\Big\}, \\
I_b &=& 2 \(1-\e\)d_A C_A g^4 \sum_f\Big\{\({\cal I}_{1}^0-\widetilde{\cal I}_{1}^0\)\widetilde{\mLARGE{\tau}}+
\fr{1}{2} \widetilde{\cal M}_{0,0}\Big\}, \\
I_c &=& \(1-\e\)d_A \(C_F - \fr{1}{2}C_A\) g^4\sum_f\Big\{4\(\I_1-2\It_1\)\Taut + \(2+\e\)\N_{0,0}
 - 2\e\Mt_{0,0} + 2\N_{1,-1}\Big\}, \\
I_d &=& -2\(1-\e\)^2 d_A C_F g^4 \sum_f \Big\{\(\I_1-\It_1\)^2\It_2-2\It_1\Taut+\Mt_{0,0}+\Mt_{1,-1}\Big\}, \\
I_e &=& -\fr{1}{2}d_A g^4\sum_{f\,g}\bigg\{4\(1+\e\)\It_1[\mu_f]\,\It_1[\mu_g]\,\I_2
+ 2\(1-\e\)\(\It_{1}[\mu_f]\,\Taut[\mu_g]+\It_{1}[\mu_g]\,\Taut[\mu_f]\)\label{pferm2} \\
&-&8\(\widetilde{\cal I}_{1}^1[\mu_f]\, \widetilde{\mLARGE{\tau}}'[\mu_g] + \widetilde{\cal I}_{1}^1[\mu_g]\,
\widetilde{\mLARGE{\tau}}'[\mu_f]\) - \(1-\e\)\N_{0,0}[\mu_f,\mu_g] - 2\N_{1,-1}[\mu_f,\mu_g] -
2\N_{2,-2}[\mu_f,\mu_g] \bigg\}, \nn
I_f &=& -\fr{1}{4}d_A C_A g^4 \sum_f \Big\{8\I_1\It_1\I_2 - 2\I_1\Taut + \Mt_{0,0} - 2\Mt_{-2,2}\Big\},
\ea
\ba
I_g &=& \fr{1}{2} d_A C_A g^4 \sum_f \Big\{4\(6-5\e\)\I_1\It_1\I_2
- \(7-6\e\)\I_1\Taut -\Big(\fr{3}{2}-2\e\Big)\Mt_{0,0}
- \(5-4\e\)\Mt_{-2,2}\Big\}, \\
I_h &=& -\(3-2\e\)\(1-\e\)d_A C_A g^4\sum_f \Big\{2\I_1\It_1\I_2-\I_1\Taut \Big\},
\label{diagb}
\ea
where the sum-integrals $\N_{n,-n}$ in Eq. (\ref{pferm2}) depend on two independent chemical potentials $\mu_f$ and $\mu_g$ through the respective fermionic momenta. The symmetry coefficients of the graphs have been taken into account here.

Using the formula
\ba
Z_g^2 \;\;\,=\;\;\, 1-\fr{11C_A-4T_F}{3}\fr{g^2}{\(4\pi\)^2}\fr{1}{\e}
\ea
for the renormalization coefficient of the gauge coupling, the unknown matching coefficient $\aE{3}$ is now available in
terms of the sum-integrals of Eqs. (\ref{ints1}) - (\ref{ints2}). One simply needs to add Eqs. (\ref{diaga}) -
(\ref{diagb}) together with the bosonic part of the strict perturbation expansion of the pressure, which can be found e.g.
from Eq. (31) of \cite{bn1}. We now turn to the actual evaluation of the unknown sum-integrals.

\section{Evaluation of the sum-integrals}
In this section all the sum-integrals encountered in appendix A will be computed. To do this we need to generalize the
results of \cite{az,zk} to finite $\mu$, which includes determining the values of some new integrals of
hyperbolic functions (see appendix C) as well as generalizing certain summation relations derived in \cite{az}. The second but last
section of this appendix deals exclusively with the problem of having different chemical potentials entering a single
sum-integral, which is the case we encounter when evaluating the diagram $I_e$. Until then it is assumed that all $\mu$'s inside
a sum-integral are equal.

\subsection{One-loop cases}
The bosonic one-loop sum-integral ${\cal I}_{n}^m$ has been evaluated in \cite{az} with the result
\ba
{\cal I}_{n}^{m} = 2^{m-2n+1}\pi^{m-2n+3/2}T^{m-2n+4}\(\fr{\Lambda^2}{\pi T^2}\)^{\e}
\fr{\Gamma(n-3/2+\e)}{\Gamma(n)}\zeta(2n-m-3+2\e).
\label{1lbos}
\ea
For the fermionic case we get after first performing a standard $3-2\e$ -dimensional integral
\ba
\widetilde{\cal I}_{n}^m
&=& \fr{1}{\(4\pi\)^{3/2-\e}}\fr{\Gamma(n-3/2+\e)}{\Gamma(n)} \Lambda^{2\e}T \sum_{k=-\infty}^{\infty}
\fr{\!\!\!\!\!\!\!\!\!\!\(\(2k+1\)\pi T-\imathb\mu\)^m}{\Big[\(\(2k+1\)\pi T-\imathb\mu\)^2\Big]^{n-3/2+\e}}  \nn
&=& 2^{m-2n}\pi^{m-2n+3/2}T^{m-2n+4}\(\fr{\Lambda^2}{\pi T^2}\)^{\e} \fr{\Gamma(n-3/2+\e)}{\Gamma(n)} \nn
&& \times \Big(\zeta(2n-m-3+2\e,1/2-\imathb \bar{\mu})+\(-1\)^m\zeta(2n-m-3+2\e,1/2+\imathb \bar{\mu})\Big),
\ea
where the definition of the generalized zeta-function
\ba
\zeta(z,q) = \sum_{n=0}^{\infty} \fr{1}{\(q+n\)^z}
\ea
has been used. The results obtained for the relevant cases after performing an $\e$-expansion are listed in Eqs.
(\ref{1loop1}) - (\ref{1loop6}).

\subsection{Two- and three-loop cases}
The computation of the two- and three-loop sum-integrals in this paper is closely analogous to the $\mu = 0$ calculations
of \cite{az} and to the two-loop work of \cite{antti} at finite $\mu$. The general scheme is to first separate the
diverging vacuum parts from the integrals and then to evaluate the rest in $d=3$. This allows a straightforward
extraction of the singularities and also simplifies the determination of the finite parts considerably. The sum-integral
$\widetilde{\mLARGE{\tau}}'$  is not considered here, as its contribution will be observed to cancel between the
different terms of Eq. (\ref{pferm2}).

\subsubsection{Preliminaries}
Let us start by deriving some results for the bosonic and fermionic `polarization' functions,
\ba
\Pi \!\(P\) &\equiv& \sumint_{Q} \fr{1}{Q^2(P+Q)^2}, \\
\Pi_{\rm f} \!\(P\) &\equiv& \sumint_{\{Q\}} \fr{1}{Q^2(P+Q)^2},
\ea
which will be frequently used in the following computations. It is straightforward to verify that in the case of a bosonic external momentum $P$ they can at $d=3$ be written in the forms
\ba
\Pi\(P\) &=& \fr{T}{\(4\pi\)^2}\int {\rm d}^3 r\,\fr{e^{\imathb \mathbf{p} \cdot \mathbf{r}}}{r^2}\( | \bar{p}_0 \! |
+ {\rm coth}\, \bar{r} \) e^{-|p_0\!| r}, \\
\Pi_{\rm f} \!\(P\) &=& \fr{T}{\(4\pi\)^2} \int {\rm d}^3 r\,\fr{e^{\imathb \mathbf{p} \cdot \mathbf{r}}}{r^2}
\( | \bar{p}_0 \!| + \cos(2\bar{\mu} \bar{r}) {\rm csch} \,\bar{r} \) e^{-|p_0\!| r},
\label{pife}
\ea
where
\ba
\bar{p}_0 &\equiv& \fr{p_0}{2\pi T}, \\
\bar{r} &\equiv& 2\pi T r.
\ea
The latter formula, Eq. (\ref{pife}), is a generalization of Eq. (4.2) of \cite{az} and has been obtained using the
Fourier-transform of the fermionic propagator,
\ba
\int\!\fr{{\rm d^{3}q}}{(2\pi)^{3}}\fr{e^{\imathb\mathbf{q}\cdot\mathbf{r}}}{\mathbf{q}^2+\(\(2n+1\)\pi T-\imathb \mu\)^2}
&=& \fr{e^{-\(|2n+1|\pi T-\imathb\mu{\rm sign}(2n+1)\)r}}{4\pi T}.
\ea

The large-$P$ behavior of the polarization functions has been analyzed in \cite{az} for the $\mu=0$ case. A straightforward generalization of the computation to non-zero chemical potentials produces for $\Pi_{\rm f}$
\ba
\label{piflp}
\Pi_{\rm f}\!\(P\) &\equiv& \Pi^{(0)}\!\(P\) + \Pi^{(T)}_{{\rm f}}\!\(P\)
\;\;\,=\;\;\, \Pi^{(0)}\!\(P\) + \Pi^{(T)}_{{\rm f,UV}}\!\(P\)+\mathcal{O}\(\fr{T^6}{P^6}\) \nn
&=& \beta_0 \(\fr{T^2}{P^2}\)^{\!\!\e}+2\widetilde{\cal I}_{1}^0\fr{1}{P^2}+\beta_2 T^4 \(\fr{1}{P^4}-\(4-2\e\)
\fr{p_0^2}{P^6}\) \\
&+& \fr{T}{\(4\pi\)^2} \!\int\! {\rm d}^3 r\,\fr{e^{\imathb \mathbf{p} \cdot \mathbf{r}}}{r^2}\!
\bigg[\cos(2\bar{\mu} \bar{r}) {\rm csch} \,\bar{r} -\fr{1}{\bar{r}}+\(\fr{1}{6}+2\bar{\mu}^2\)\bar{r}-
\(\fr{7}{360}+\fr{\bar{\mu}^2}{3}+\fr{2\bar{\mu}^4}{3}\)\bar{r}^3\bigg]e^{-|p_0\!| r}, \nonumber
\ea
where $\e$ has been set to zero in the last $\mathcal{O}\!\(\fr{T^6}{P^6}\!\)$-term. Here the coefficients $\beta$ read
\ba
\beta_0 &=& \fr{1}{\(4\pi\)^{2-\e}}\fr{\Gamma (\e)\Gamma^2(1-\e)}{\Gamma (2-2\e)}\(\fr{\Lambda}{T}\)^{\!2\e}, \\
\beta _2 &=& 2^{2\e}\pi^{-3/2+\e}\fr{\Gamma (3-2\e )}{\Gamma (3/2-\e )}\(\fr{\Lambda}{T}\)^{\!2\e}
\( {\rm Li}_{4-2\e}(\!-e^{\mu /T})+{\rm Li}_{4-2\e}(\!-e^{-\mu /T})\)
\ea
with Li denoting the polylogarithm function, and the relation
\ba
{\rm Li}_{4}(-e^{x}) + {\rm Li}_{4}(-e^{-x}) = -\fr{7\pi^4}{360}-\fr{\pi^2}{12}x^2-\fr{1}{24}x^4
\ea
has been employed in deriving the last term of Eq. (\ref{piflp}). The first term of Eq. (\ref{piflp}), denoted by $\Pi^{(0)}$, is the vacuum ($T=\mu=0$) part of the polarization function, and is the same in the bosonic and fermionic cases. Following the notation of \cite{az}, the vacuum subtracted, or matter, part of the function is denoted by $\Pi^{(T)}_{{\rm f}}$, even though here it should not be confused with the finite temperature part of $\Pi_{\rm f}$ obtained simply by subtracting the $T=0$ piece from the polarization function.

\subsubsection{$\widetilde{\mLARGE{\tau}}$}
In terms of the function $\Pi_{\rm f}$ and its large-$P$ expansion, the sum-integral $\widetilde{\mLARGE{\tau}}$ can clearly be written as
\ba
\widetilde{\mLARGE{\tau}} &=& \sumint_P \fr{1}{P^2}\Pi_{\rm f}\!\(P\) \;\;\,=\;\;\, \sumint_P \fr{1}{P^2}
\bigg( \beta_0 \(\fr{T^2}{P^2}\)^{\!\!\e}+2\widetilde{\cal I}_{1}^0\fr{1}{P^2} \nn
&+& \fr{T}{\(4\pi\)^2} \!\int\! {\rm d}^3 r\,\fr{e^{\imathb \mathbf{p} \cdot \mathbf{r}}}{r^2}
\bigg[\cos(2\bar{\mu} \bar{r}) {\rm csch} \,\bar{r} -\fr{1}{\bar{r}}
+ \(\fr{1}{6}+2\bar{\mu}^2\)\bar{r}\bigg]e^{-|p_0\!| r}\bigg) + \mathcal{O}\(\e\).
\label{tau1}
\ea
The first two terms are trivial to evaluate using Eq. (\ref{1lbos}), whereas the last one gives at $d = 3$
\ba
&& \fr{T}{\(4\pi\)^2} \sumint_P \fr{1}{P^2}\!\int\! {\rm d}^3 r\,\fr{e^{\imathb \mathbf{p} \cdot \mathbf{r}}}{r^2}
\bigg[\cos(2\bar{\mu} \bar{r}) {\rm csch} \,\bar{r} -\fr{1}{\bar{r}}
+ \(\fr{1}{6}+2\bar{\mu}^2\)\bar{r}\bigg]e^{-|p_0\!| r} \nn
&=& \fr{T^2}{\(4\pi\)^3}\!\int\! {\rm d}^3 r \sum_{p_0} \fr{e^{-2|p_0\!|r}}{r^3}
\bigg[\cos(2\bar{\mu} \bar{r}) {\rm csch} \,\bar{r} -\fr{1}{\bar{r}}
+ \(\fr{1}{6}+2\bar{\mu}^2\)\bar{r}\bigg] \nn
&=& \fr{T^2}{\(4\pi\)^2}\int_0^{\infty} {\rm d} r \,\fr{{\rm coth }\,r}{r}
\bigg[\cos(2\bar{\mu} r) {\rm csch} \,r -\fr{1}{r} + \(\fr{1}{6}+2\bar{\mu}^2\)r\bigg].
\label{tau2}
\ea
Here the final integral is UV convergent but has an IR divergence due to the zero mode of the bosonic frequency sum. Using the
results of appendix C it can, however, be evaluated analytically and produces a finite result. This is due
to the fact that the divergence in the $r$-integral comes from a term of the form $\int_0^{\infty}{\rm d}r \, r^{\alpha}$,
which vanishes under dimensional regularization. Adding up the different pieces and expanding in $\e$, we get the result, Eq. (\ref{taub}), for the whole sum-integral.

\subsubsection{$\N_{0,0}$ and $\Mt_{0,0}$}
The application of Eq. (\ref{piflp}) separates the evaluation of $\N_{0,0}$ into three pieces
\ba
\N_{0,0} \;=\; \sumint_P \(\Pi^{(T)}_{{\rm f}}\!\(P\)\)^2 \;=\; \sumint_P \(\(\Pi^{(0)}\!\(P\)\)^2+\(\Pi^{(T)}_{{\rm f}}\!\(P\)\)^2+
2\,\Pi^{(0)}\!\(P\)\Pi^{(T)}_{{\rm f}}\!\(P\)\),
\label{n001}
\ea
of which the first one is again trivial. The second produces in analogy with the above two-loop calculation
\ba
&& \sumint_P \Pi^{(T)}_{{\rm f}}\!\(P\)^2 \;\;\,=\;\;\, \sumint_P \(\(\Pi^{(T)}_{{\rm f}}\!\(P\)\)^2
-4\(\widetilde{\cal I}_{1}^0\)^2\fr{1}{P^4}\)+4\(\widetilde{\cal I}_{1}^0\)^2{\cal I}_{2}^0 \nn
&=& \fr{T^3}{\(4\pi\)^4}\int {\rm d}^3 r \sum_{p_0} \fr{e^{-2|p_0\!|r}}{r^4}
\bigg[\(\cos(2\bar{\mu} \bar{r}) {\rm csch} \,\bar{r} -\fr{1}{\bar{r}}\)^2- \(\fr{1}{6}+2\bar{\mu}^2\)^2\bar{r}^2\bigg]
+ 4\(\widetilde{\cal I}_{1}^0\)^2{\cal I}_{2}^0 \nn
&=& \fr{T^4}{32\pi^2}\int_0^{\infty} {\rm d}r\, \fr{{\rm coth }\,r}{r^2}
\bigg[\(\cos(2\bar{\mu} r) {\rm csch} \,r -\fr{1}{r}\)^2- \(\fr{1}{6}+2\bar{\mu}^2\)^2r^2\bigg]
+ 4\(\widetilde{\cal I}_{1}^0\)^2{\cal I}_{2}^0,
\ea
where the resulting one-dimensional integral is analytically calculable.

The third term of $\N_{0,0}$ can be divided into three parts
\ba
\sumint_P \Pi^{(0)}\!\(P\)\Pi^{(T)}_{{\rm f}}\!\(P\) &=& \sumint_P\(\Pi^{(0)}\!\(P\)-
\fr{1}{\(4\pi\)^2\e}\)\(\Pi^{(T)}_{{\rm f}}\!\(P\)-\Pi^{(T)}_{{\rm f,UV}}\!\(P\)\) \\
&+&\fr{1}{\(4\pi\)^2\e}\sumint_P\(\Pi^{(T)}_{{\rm f}}\!\(P\)-\Pi^{(T)}_{{\rm f,UV}}\!\(P\)\)
+\sumint_P \Pi^{(0)}\!\(P\)\Pi^{(T)}_{{\rm f,UV}}\!\(P\) \nn
&\equiv& K_1 + K_2 + K_3.
\label{pip0m00}
\ea
Here $K_2$ and $K_3$ are straightforward to obtain using the previous results, but the first term requires careful consideration. Taking into account
that
\ba
\sumint_P\(\Pi^{(T)}_{{\rm f}}\!\(P\)-\Pi^{(T)}_{{\rm f,UV}}\!\(P\)\) &=& \mathcal{O}\(\e\)
\ea
and using the result of \cite{az},
\ba
\int\! \fr{{\rm d}^3 p}{\(2\pi\)^3}e^{\imathb\mathbf{p}\cdot\mathbf{r}}\,\ln \,\fr{4\pi \Lambda^2}{\mathbf{p}^2+p_0^2}
&=& \fr{1}{2\pi r}\(\fr{|p_0\!|}{r}+\fr{1}{r^2}\)e^{-|p_0\!|r},
\label{lnint}
\ea
we get
\ba
K_1 &=& \fr{T}{\(4\pi\)^4}\sumint_P \bigg(\ln \,\fr{4\pi \Lambda^2}{P^2}+2-\gamma\bigg)\!
\int\! {\rm d}^3 r\fr{e^{\imathb \mathbf{p} \cdot \mathbf{r}}}{r^2}
\bigg[\cos(2\bar{\mu} \bar{r}) {\rm csch} \,\bar{r} -\fr{1}{\bar{r}} + \(\fr{1}{6}+2\bar{\mu}^2\)\bar{r} \nn
&-& \(\fr{7}{360}+\fr{\bar{\mu}^2}{3}+\fr{2\bar{\mu}^4}{3}\)\bar{r}^3\bigg]e^{-|p_0\!| r} \nn
&=& \fr{T^4}{32\pi^2}\int_0^{\infty}\!{\rm d}r \, r^{-3}\bigg[\cos(2\bar{\mu} r) {\rm csch} \,r -\fr{1}{r}
+ \(\fr{1}{6}+2\bar{\mu}^2\)r \nn
&-& \(\fr{7}{360}+\fr{\bar{\mu}^2}{3}+\fr{2\bar{\mu}^4}{3}\)r^3\bigg]
\(1-\fr{r}{2}\fr{{\rm d}}{{\rm d}r}\){\rm coth}\,r.
\ea
The remaining integral is again of analytically solvable type, and after adding together all the different pieces we get as the result Eq. (\ref{n00res}).

The computation of $\Mt_{0,0}$ proceeds in a similar fashion. Eq. (\ref{n001}) is replaced by
\ba
\Mt_{0,0} &=& \sumint_P \(\(\Pi^{(0)}\!\(P\)\)^2+\Pi^{(T)}\!\(P\)\Pi^{(T)}_{{\rm f}}\!\(P\) +
\Pi^{(0)}\!\(P\)\(\Pi^{(T)}\!\(P\)+\Pi^{(T)}_{{\rm f}}\!\(P\)\)\),
\label{mt001}
\ea
where the first term is trivial and the last two cross-terms are available through Eq. (\ref{pip0m00}) above and Eq. (D20) of
\cite{az}. For the remaining piece one obtains
\ba
\label{m00int1}
&&\sumint_P \Pi^{(T)}\!\(P\) \Pi^{(T)}_{{\rm f}}\!\(P\) \;\;=\;\;
\sumint_P \(\Pi^{(T)}\!\(P\) \Pi^{(T)}_{{\rm f}}\!\(P\)
-4 {\cal I}_{1}^0 \widetilde{\cal I}_{1}^0\fr{1}{P^4}\)+
4{\cal I}_{1}^0\widetilde{\cal I}_{1}^0{\cal I}_{2}^0 \\
&=& \fr{T^4}{32\pi^2}\int_0^{\infty} {\rm d} r \,\fr{{\rm coth }\,r}{r^2}
\bigg[\({\rm coth}\,r-\fr{1}{r}\)\(\cos(2\bar{\mu} r) {\rm csch} \,r -\fr{1}{r}\) +
\(\fr{1}{18}+\fr{2\bar{\mu}^2}{3}\)r^2\bigg] + 4{\cal I}_{1}^0\widetilde{\cal I}_{1}^0{\cal I}_{2}^0, \nonumber
\ea
whose straightforward evaluation leads to the final result, Eq. (\ref{m00res1}).

\subsubsection{$\N_{1,-1}$ and $\Mt_{1,-1}$}
It is easy to see that the sum-integrals $\N_{1,-1}$ and $\Mt_{1,-1}$ can be written in the form
\ba
\N_{1,-1} &=& 2 \widetilde{\cal I}_{1}^0 \widetilde{\mLARGE{\tau}}
- 2\sumint_{P\{QR\}}\fr{Q\cdot R}{P^2Q^2R^2(P+Q)^2(P+R)^2} \;\;\,\equiv\;\;\,
2 \widetilde{\cal I}_{1}^0 \widetilde{\mLARGE{\tau}} - 2{\rm H_{3}'}, \\
\Mt_{1,-1} &=& 2 \widetilde{\cal I}_{1}^0 \widetilde{\mLARGE{\tau}}
- 2\sumint_{\{P\}QR}\fr{Q\cdot R}{P^2Q^2R^2(P+Q)^2(P+R)^2} \;\;\,\equiv\;\;\,
2 \widetilde{\cal I}_{1}^0 \widetilde{\mLARGE{\tau}} - 2{\rm H_{3}}´.
\ea
Following \cite{az} and defining further
\ba
J'_{\mu}\!\(P\) &\equiv& \sumint_{\{Q\}} \fr{\(2Q+P\)_{\mu}}{Q^2\(Q+P\)^2}, \\
J_{\mu}\!\(P\) &\equiv& \sumint_{Q} \fr{\(2Q+P\)_{\mu}}{Q^2\(Q+P\)^2}-\fr{P_{\mu}}{P^2}\({\cal I}_{1}^0-
\widetilde{\cal I}_{1}^0\), \\
\label{i3'}
{\rm I_3'} &\equiv& \sumint_P \fr{1}{P^2}J'_{\mu}\!\(P\)J'_{\mu}\!\(P\), \\
{\rm I_3} &\equiv& \sumint_{\{P\}} \fr{1}{P^2}J_{\mu}\!\(P\)J_{\mu}\!\(P\),
\label{i3}
\ea
we quickly verify that ${\rm H_{3}'}$ and ${\rm H_{3}}$ read
\ba
{\rm H_{3}'} &=& \fr{1}{4}{\rm I_3'}+\fr{1}{4}\N_{0,0}, \\
{\rm H_{3}} &=& \fr{1}{4}{\rm I_3}+\fr{1}{4}\Mt_{0,0}+\fr{1}{4}\({\cal I}_{1}^0 -
\widetilde{\cal I}_{1}^0\)^2\widetilde{\cal I}_{2}^0-\fr{1}{2}\({\cal I}_{1}^0 -
\widetilde{\cal I}_{1}^0\)\widetilde{\mLARGE{\tau}}.
\ea
This leaves only the simpler sum-integrals ${\rm I_{3}'}$ and ${\rm I_{3}}$ to be evaluated.

Due to Lorentz invariance and the orthogonality of $J'_{\mu}\!\(P\)$ to bosonic $P$ and $J_{\mu}\!\(P\)$ to fermionic $P$
it is evident that these functions can be written in the forms
\ba
J'_{\mu}\!\(P\) &=& \fr{P^2}{p^2}\(\delta_{\mu,0}-\fr{p_0}{P^2}P_{\mu}\)j'_0\!\(P\), \\
J_{\mu}\!\(P\) &=& \fr{P^2}{p^2}\(\delta_{\mu,0}-\fr{p_0}{P^2}P_{\mu}\)j_0\!\(P\),
\ea
which, when plugged into Eqs. (\ref{i3'}) and (\ref{i3}), produce
\ba
\label{i3'2}
{\rm I_3'} &=& \sumint_P \fr{1}{p^2}j'_{0}\!\(P\)^2, \\
{\rm I_3} &=& \sumint_{\{P\}} \fr{1}{p^2}j_{0}\!\(P\)^2.
\label{i32}
\ea
A crucial simplification in the evaluation of these sum-integrals occurs, as one notices that they both are actually finite. This is due to the fact that at large $P$ $J'_{\mu}\!\(P\)$ and $J_{\mu}\!\(P\)$ behave like $\mathcal{O}\(1/P^2\)$, as can be straightforwardly verified. We may therefore set $\e = 0$ in the expressions for $j'_{0}$ and $j_{0}$, which eventually yields
\ba
\label{j0'}
j'_{0}\!\(P\) &=& -\fr{\imathb T}{4\pi}\!\int_0^{\infty}\!{\rm d} r\(\partial_r \fr{\sin \,pr}{pr}\)
\Big[2\mubar-\sin(2\mubar \bar{r}){\rm csch}\,\bar{r}\Big]e^{-|p_0\!|r}, \\
j_{0}\!\(P\) &=& -\fr{T}{8\pi}\!\int_0^{\infty}\!{\rm d} r\(\partial_r \fr{\sin \,pr}{pr}\)
\bigg[e^{-2\imathb \mubar \bar{r}\, {\rm sign}({\rm Re}\,p_0)}{\rm csch}\,\bar{r}-{\rm coth}\,\bar{r} \nn
&& + \fr{1}{2}\Big(\(1+4\mubar^2\)\bar{r}+4\imathb\mubar\,{\rm sign}({\rm Re}\,p_0)\Big)\bigg]
e^{-\(|{\rm Re}(p_0)|-\imathb\mu\,{\rm sign}({\rm Re}\,p_0)\)r}\,{\rm sign}({\rm Re}\,p_0).
\label{j0}
\ea

After substituting Eqs. (\ref{j0'}) and (\ref{j0}) to Eqs. (\ref{i3'2}) and (\ref{i32}), doing the $p_0$-sums and performing the remaining $p$-integrals using the relation
\ba
\int\fr{{\rm d}^3 p}{\(2\pi\)^3}\fr{1}{p^2}\fr{\sin\,pr}{pr}\fr{\sin\,ps}{ps} &=& \fr{1}{4\pi}\Big(\fr{1}{r}\theta (r-s)
-\fr{1}{s}\theta (s-r)\Big),
\ea
we finally get
\ba
{\rm I_{3}'} &=& -\fr{T^4}{32\pi^2}\int_0^{\infty}\!{\rm d}r\, \fr{{\rm coth}\,r}{r^2}
\Big(2\mubar-\sin(2\mubar r){\rm csch}\,r\Big)^2, \\
{\rm I_{3}} &=& \fr{T^4}{128\pi^2}{\rm Re}\Bigg\{\int_0^{\infty}\!{\rm d}r\, \fr{e^{2\imathb \mubar r}{\rm csch}\,r}{r^2}
\bigg(e^{-2\imathb \mubar r}{\rm csch}\,r-{\rm coth}\,r
+ \fr{1}{2}\Big(\(1+4\mubar^2\)r+4\imathb\mubar\Big)\bigg)^2\Bigg\}.
\ea
These integrals are clearly both UV- and IR-finite and are readily evaluated. The final results for the sum-integrals, obtained after adding up all the different pieces, are given by Eqs. (\ref{n11res}) and (\ref{m11res}).

\subsubsection{$\N_{2,-2}$ and $\Mt_{-2,2}$}
Let us define, once again in analogy with \cite{az}, the modified gluon polarization tensors,
\ba
\bar{\Pi}_{\mu\nu}\!\(P\) &\equiv& 2\delta_{\mu\nu}\I_1 - \sumint_{Q}\fr{\(2Q+P\)_{\mu}\(2Q+P\)_{\nu}}{Q^2\(Q+P\)^2},\\
\bar{\Pi}_{{\rm f},\mu\nu}\!\(P\) &\equiv& 2\delta_{\mu\nu}\It_1 -
\sumint_{\{Q\}}\fr{\(2Q+P\)_{\mu}\(2Q+P\)_{\nu}}{Q^2\(Q+P\)^2}.
\ea
Using these we further define two new sum-integrals by
\ba
I_{sqed}^{ff} &\equiv& \sumint_P\fr{1}{P^4}\(\Delta\bar{\Pi}_{{\rm f},\mu\nu}\!\(P\)\)^2, \\
I_{sqed}^{bf} &\equiv& \sumint_P\fr{1}{P^4}\Delta\bar{\Pi}_{\mu\nu}\!\(P\)\Delta\bar{\Pi}_{{\rm f},\mu\nu}\!\(P\),
\ea
where $\Delta f\(P\) \equiv f\(P\) - f\(0\)\!\delta_{p_0,0}$. It is
then straightforward to verify that determining $\N_{2,-2}$ and $\Mt_{-2,2}$ can be reduced to computing a set of simpler sum-integrals
\ba
\label{n22exp}
\N_{2,-2} &=& \fr{1}{4}I_{sqed}^{ff} + \fr{1}{2}\Delta I_{sqed}^{ff} - \N_{1,-1} -\fr{1}{4}\N_{0,0} -
8\widetilde{\cal I}_{1}^1 \widetilde{\mLARGE{\tau}}' + \(2+2\e\)\(\It_1\)^2\I_2, \label{M2} \\
\Mt_{-2,2} &=& \fr{1}{4}I_{sqed}^{bf}+\fr{1}{4}\Delta I_{sqed}^{bf}+\fr{1}{4}\Mt_{0,0} -\I_1\Taut+\(2+2\e\)\I_1\It_1\I_2,
\ea
where
\ba
\Delta I_{sqed}^{ff} &\equiv& T\int \! \fr{{\rm d}^{3-2\e}p}{\(2\pi\)^{3-2\e}}
\fr{1}{p^4}\bar{\Pi}_{{\rm f},\mu\nu}\(0\)\bar{\Pi}_{{\rm f},\mu\nu}\(p_0=0,p\), \label{dff} \\
\Delta I_{sqed}^{bf} &\equiv& T\int \! \fr{{\rm d}^{3-2\e}p}{\(2\pi\)^{3-2\e}}\fr{1}{p^4}
\Big[\bar{\Pi}_{\mu\nu}\(0\)\bar{\Pi}_{{\rm f},\mu\nu}\(p_0=0,p\) +
\bar{\Pi}_{{\rm f},\mu\nu}\(0\)\bar{\Pi}_{\mu\nu}\(p_0=0,p\)\Big]. \label{dbf}
\ea

The evaluation of $I_{sqed}^{ff}$ and $I_{sqed}^{bf}$ is fairly easy, since an elementary calculation verifies
the validity of Eqs. (F6) - (F10) of \cite{az} also when $\mu \neq 0$. This gives at order $\mathcal{O}\(\e\)$
\ba
\label{n22ens}
\sumint_P\fr{1}{P^4}\(\Delta\bar{\Pi}_{{\rm f},\mu\nu}^{(T)}\!\(P\)\)^2 &=& \sumint_P \(\Pi^{(T)}_{{\rm f}}\!\(P\)\)^2
+ 4\(d-2\)\(\It_1\)^2\I_2,
\ea
\ba
\sumint_P\fr{1}{P^4}\Delta\bar{\Pi}_{\mu\nu}^{(T)}\!\(P\)\Delta\bar{\Pi}_{{\rm f},\mu\nu}^{(T)}\!\(P\)
&=& \sumint_P \Pi^{(T)}\!\(P\)\Pi^{(T)}_{{\rm f}}\!\(P\) + 4\(d-2\)\I_1 \It_1 \I_2, \\
\sumint_P\fr{1}{P^4}\Delta\bar{\Pi}_{{\rm f},\mu\nu}^{(T)}\!\(P\) \bar{\Pi}_{\mu\nu}^{(0)}\!\(P\) &=&
\fr{1}{d-1}\sumint_P \Pi^{(T)}_{{\rm f}}\!\(P\)\Pi^{(0)}\!\(P\)+
\fr{2\(d-2\)}{d-1}\It_1\sumint_P\fr{1}{P^2}\Pi^{(0)}\!\(P\), \\
\sumint_P\fr{1}{P^4}\Delta\bar{\Pi}_{\mu\nu}^{(T)}\!\(P\) \bar{\Pi}_{\mu\nu}^{(0)}\!\(P\) &=&
\fr{1}{d-1}\sumint_P \Pi^{(T)}\!\(P\)\Pi^{(0)}\!\(P\)+\fr{2\(d-2\)}{d-1}\I_1\sumint_P\fr{1}{P^2}\Pi^{(0)}\!\(P\), \\
\sumint_P\fr{1}{P^4}\(\bar{\Pi}_{\mu\nu}^{(0)}\!\(P\)\)^2 &=& \fr{1}{d-1}\sumint_P\(\Pi^{(0)}\!\(P\)\)^2,
\label{n22b}
\ea
where the superscript $(0)$ again signifies the vacuum part of the function in question and $f^{(T)}\equiv f-f^{(0)}$.

The relations (\ref{n22ens}) - (\ref{n22b}) clearly reduce $I_{sqed}^{ff}$ and $I_{sqed}^{bf}$ to well-known sum-integrals and leave only $\Delta I_{sqed}^{ff}$ and $\Delta I_{sqed}^{bf}$ to be evaluated. Noticing that for $P = (0,p)$
\ba
P_{\mu}\bar{\Pi}_{{\rm f},\mu\nu}\(0\) &=&
2P_{\nu}\It_1-\fr{4 P_{\nu}}{3-2\e}\sumint_{\{Q\}}\fr{Q^2-q_{0}^{2}}{Q^4} \nn
&=&\fr{P_{\nu}}{3-2\e}\(\(2-4\e\)\It_1+4\widetilde{\cal I}_{2}^2\) \nn
&=& 0,
\ea
and similarly
\ba
P_{\mu}\bar{\Pi}_{\mu\nu}\(0\) \;\;=\;\; P_{\mu}\bar{\Pi}_{{\rm f},\mu\nu}\(0,p\)
\;\;=\;\; P_{\mu}\bar{\Pi}_{\mu\nu}\(0,p\)
\;\;=\;\; 0,
\ea
we may decompose the tensors into their transverse and longitudinal parts (see e.g. \cite{kap}). This enables us to write Eqs.
(\ref{dff}) and (\ref{dbf}) in the form
\ba
\Delta I_{sqed}^{ff} &=& T\int \! \fr{{\rm d}^{3-2\e}p}{\(2\pi\)^{3-2\e}}
\fr{1}{p^4}\bigg[\fr{3-2\e}{2-2\e}\bar{\Pi}_{{\rm f},00}\(0\)\bar{\Pi}_{{\rm f},00}\(0,p\) \nn
&-&\fr{1}{2-2\e}\Big(\bar{\Pi}_{{\rm f},00}\(0\)\bar{\Pi}_{{\rm f},\mu\mu}\(0,p\) +
\bar{\Pi}_{{\rm f},\mu\mu}\(0\)\bar{\Pi}_{{\rm f},00}\(0,p\)\Big)
+ \fr{1}{2-2\e}\bar{\Pi}_{{\rm f},\mu\mu}\(0\)\bar{\Pi}_{{\rm f},\mu\mu}\(0,p\)\bigg], \label{dff2}\\
\Delta I_{sqed}^{bf} &=& T\int \! \fr{{\rm d}^{3-2\e}p}{\(2\pi\)^{3-2\e}}
\fr{1}{p^4}\bigg[\fr{3-2\e}{2-2\e}\(\bar{\Pi}_{{\rm f},00}\(0\)\bar{\Pi}_{00}\(0,p\) +
\bar{\Pi}_{00}\(0\)\bar{\Pi}_{{\rm f},00}\(0,p\)\) \nn
&-&\fr{1}{2-2\e}\Big(\bar{\Pi}_{{\rm f},00}\(0\)\bar{\Pi}_{\mu\mu}\(0,p\) +
\bar{\Pi}_{{\rm f},\mu\mu}\(0\)\bar{\Pi}_{00}\(0,p\)
+\bar{\Pi}_{00}\(0\)\bar{\Pi}_{{\rm f},\mu\mu}\(0,p\) +
\bar{\Pi}_{\mu\mu}\(0\)\bar{\Pi}_{{\rm f},00}\(0,p\)\Big) \nn
&+& \fr{1}{2-2\e}\Big(\bar{\Pi}_{{\rm f},\mu\mu}\(0\)\bar{\Pi}_{\mu\mu}\(0,p\) +
\bar{\Pi}_{\mu\mu}\(0\)\bar{\Pi}_{{\rm f},\mu\mu}\(0,p\)\Big)\bigg], \label{dbf2}
\ea
and the use of the identity $\widetilde{\cal I}_{2}^2 / \It_1 = {\cal I}_{2}^2 / \I_1 = -1/2+\e$ now leads to
\ba
\Delta I_{sqed}^{ff} &=& T\int \! \fr{{\rm d}^{3-2\e}p}{\(2\pi\)^{3-2\e}}\fr{1}{p^4}
\bigg[\fr{2}{2-2\e}\(\(1-2\e\)\It_1 + 2\widetilde{\cal I}_{2}^2\)p^2\bar{\Pi}_{{\rm f}}\(0,p\) \nn
&+&\fr{8}{2-2\e}\(-\It_1+2\(3-2\e\)\widetilde{\cal I}_{2}^2\)\sumint_{\{Q\}}\fr{q_0^2}{Q^2\(Q+p\)^2}\bigg] \nn
&=&-8\(2-2\e\)T\It_1 \sumint_{\{Q\}}\int \! \fr{{\rm d}^{3-2\e}p}{\(2\pi\)^{3-2\e}}\fr{q_0^2}{p^4Q^2(Q+p)^2}, \label{jnk}\\
\Delta I_{sqed}^{bf} &=&
-8\(2-2\e\)T\bigg(\It_1\sumint_{\{Q\}}\!
+\,\I_1\sumint_{Q}\bigg)\int\!\fr{{\rm d}^{3-2\e}p}{\(2\pi\)^{3-2\e}}\fr{q_0^2}{p^4Q^2(Q+p)^2}. \label{hjk}
\ea

In Eqs. (\ref{jnk}) and (\ref{hjk}) the $p$-integrals can be performed by introducing Feynman parameters $x$ and $y$ to combine the different factors in the denominators. A straightforward calculation produces
\ba
\label{dff5}
\Delta I_{sqed}^{ff} &=& -\fr{16\(1-\e\)\Gamma(5/2+\e)}{\(4\pi\)^{3/2-\e}}T\,\It_1\,
\widetilde{\cal I}_{5/2+\e}^2\,{\rm f}(\e), \\
\Delta I_{sqed}^{bf} &=& -\fr{16\(1-\e\)\Gamma(5/2+\e)}{\(4\pi\)^{3/2-\e}}T
\(\I_1\,\widetilde{\cal I}_{5/2+\e}^2+\It_1\,{\cal I}_{5/2+\e}^2\){\rm f}(\e), \label{dbf5}
\ea
where the function $\rm{f}$ is defined by
\ba
{\rm f}(\e) &\equiv& \int_0^1{\rm d}x\int_0^1{\rm d}y \, x^{-3/2-\e}\(1-x\)y^{1/2+\e}\(1-xy\)^{-3/2+\e} \nn
&=&\int_0^1{\rm d}x\int_0^1{\rm d}y \, x^{-3/2}\(1-x\)y^{1/2}\(\(1-xy\)^{-3/2}-1-\fr{3xy}{2}\) \nn
&+&\int_0^1{\rm d}x\int_0^1{\rm d}y \, x^{-3/2-\e}\(1-x\)y^{1/2+\e}\(1+\fr{3xy}{2}\) + \mathcal{O}\(\e\) \nn
&=& -\fr{\pi}{2} + \mathcal{O}\(\e\).
\label{fe}
\ea
The use of this result in Eqs. (\ref{dff5}) and (\ref{dbf5}) now gives at order $\mathcal{O}\(\e\)$
\ba
\Delta I_{sqed}^{ff} &=& 0, \\
\Delta I_{sqed}^{bf} &=& \fr{T^4}{12\(4\pi\)^2}\(1+12\mubar^2\),
\ea
and together with the other parts of $\N_{2,-2}$ and $\Mt_{-2,2}$ this yields Eqs. (\ref{n22resa}) and (\ref{m22resa}) as the results.

\subsubsection{The diagram $e$}
Eq. (\ref{pferm2}), the contribution of the diagram $e$ to the strict perturbation expansion of the QCD pressure,
contains sum-integrals, which are functions of two independent chemical potentials. Due to cancellations between the
different terms of $I_e$ it is convenient not to deal with all of them separately, but to treat the whole diagram as a single entity.

Using trivial generalizations of Eqs. (\ref{n22exp}) and (\ref{n22ens}) we get for the diagram
\ba
I_e &=& -\fr{1}{2}d_A g^4 \sum_{f\,g}\bigg\{4\(1-\e\)\It_1(\mu_f)\,\It_1(\mu_g)\,\I_2
- 2\(1-\e\)\(\It_{1}(\mu_f)\,\Taut(\mu_g)+\It_{1}(\mu_g)\,\Taut(\mu_f)\) \nn
&+& \(1-\e\)\sumint_P \Pi^{(T)}_{{\rm f}}\!\(P,\mu_f\)\Pi^{(T)}_{{\rm f}}\!\(P,\mu_g\)
+\fr{2\(1-\e\)}{3-2\e}\(\It_1(\mu_f)+\It_1(\mu_g)\)\sumint_P\fr{1}{P^2}\Pi^{(0)}\!\(P\)\nn
&+& \fr{2\(1-\e\)^2}{3-2\e}\sumint_P\Pi^{(0)}\!\(P\)\(\Pi^{(0)}\!\(P\)+
\Pi^{(T)}_{{\rm f}}\!\(P,\mu_f\) + \Pi^{(T)}_{{\rm f}}\!\(P,\mu_g\)\) \bigg\} \nn
&=& -\fr{1}{4}d_A g^4 \sum_{f\,g}\bigg\{16\(1-\e\)\It_1(\mu_f)\,\It_1(\mu_g)\,\I_2
- 4\(1-\e\)\(\It_{1}(\mu_f)\,\Taut(\mu_g)+\It_{1}(\mu_g)\,\Taut(\mu_f)\) \nn
&+& \fr{\(1-\e\)T^4}{\(4\pi\)^2}\int_0^{\infty} {\rm d}r\, \fr{{\rm coth }\,r}{r^2}
\bigg[\(\cos(2\bar{\mu}_f r) {\rm csch} \,r -\fr{1}{r}\)\(\cos(2\bar{\mu}_g r) {\rm csch} \,r -\fr{1}{r}\) \nn
&-& \(\fr{1}{6}+2\bar{\mu}_f^2\)\(\fr{1}{6}+2\bar{\mu}_g^2\)r^2\bigg]
+ \fr{4\(1-\e\)^2}{3-2\e}\bigg[\sumint_P\Pi^{(0)}\!\(P\)\(\Pi^{(0)}\!\(P\)+
\Pi^{(T)}_{{\rm f}}\!\(P,\mu_f\) + \Pi^{(T)}_{{\rm f}}\!\(P,\mu_g\)\) \nn
&+&\fr{\It_1(\mu_f) + \It_1(\mu_g)}{1-\e}\sumint_P\fr{1}{P^2}\Pi^{(0)}\!\(P\)\bigg] \bigg\},
\ea
where every term is of an already familiar form with the exception of the one containing the $r$-integral. This
integral can, however, also be straightforwardly evaluated using the results of appendix C, which eventually gives us Eq.
(\ref{mufmugres}) as the final outcome of the graph.

\subsection{The results}
The final results for the sum-integrals introduced in appendix A  and evaluated above read
\ba
{\cal I}_{1}^{0}&=&\fr{T^2}{12}\(1+2\e\bigg[1+\fr{\zeta'(-1)}{\zeta(-1)}+\ln\,\fr{\bar{\Lambda}}{4\pi T}\bigg]\), \label{1loop1}\\
\widetilde{\cal I}_{1}^0 &=& -\fr{T^2}{24}\bigg(1+12\mubar^2
+2\e\bigg[\(1+12\mubar^2\)\(1+\ln\,\fr{\bar{\Lambda}}{4\pi T}\)
+12\,\aleph(1,z)\bigg]\bigg), \\
{\cal I}_{2}^{0} &=& \fr{1}{\(4\pi\)^2}\(\fr{1}{\e}+2\gamma+2\,\ln\,\fr{\bar{\Lambda}}{4\pi T}\), \\
\widetilde{\cal I}_{2}^0 &=& \fr{1}{\(4\pi\)^2}\(\fr{1}{\e}-\aleph(z) +
2\,\ln\,\fr{\bar{\Lambda}}{4\pi T}\), \\
\widetilde{\cal I}_{2}^2 &=& \fr{T^2}{48}\bigg(1+12\mubar^2
+2\e\bigg[\(1+12\mubar^2\)\ln\,\fr{\bar{\Lambda}}{4\pi T}
+12\,\aleph(1,z)\bigg]\bigg), \\
\label{1loop6}
\widetilde{\mLARGE{\tau}} &=& -\fr{T^2}{\(4\pi\)^2}\(\fr{\bar{\mu}^2}{\e} +
2\mubar^2\(1+2\,\ln\,\fr{\bar{\Lambda}}{4\pi T}\) -
2\imathb \mubar \aleph(0,z)\),
\label{taub} \\
\N_{0,0} &=& \fr{1}{\(4\pi\)^2}\(\fr{T^2}{12}\)^2\bigg(\fr{3}{2}\(1+12\mubar^2\)^2\(\fr{1}{\e}+
6\,\ln\,\fr{\bar{\Lambda}}{4\pi T}\) +\fr{173}{20}
+ 210\mubar^2 + 1284\mubar^4 -\fr{24}{5}\fr{\zeta'(-3)}{\zeta(-3)} \nn
&-& 144\Big[2\aleph(3,z) + \aleph(3,2z) + 4\imathb\mubar\Big(\aleph(2,z) + \aleph(2,2z)\Big) - \(1+8\mubar^2\)\aleph(1,z)
- 4\mubar^2\aleph(1,2z)\Big]\bigg)\!,
\label{n00res} \\
\Mt_{0,0} &=& -\fr{1}{\(4\pi\)^2}\(\fr{T^2}{12}\)^2\bigg(\fr{3}{4}\(1+24\mubar^2-48\mubar^4\)\(\fr{1}{\e}+
6\,\ln\,\fr{\bar{\Lambda}}{4\pi T}\) + \fr{179}{40} + 111\mubar^2 - 210\mubar^4 \nn
&+& 48\fr{\zeta '(-1)}{\zeta (-1)} + \fr{24}{5}\fr{\zeta '(-3)}{\zeta (-3)}
+ 72\Big[6\aleph(3,z) + 12\imathb\mubar\,\aleph(2,z) + \(1-8\mubar^2\)\aleph(1,z)\Big]\bigg), \label{m00res1} \\
\label{n11res} \N_{1,-1} &=& -\fr{1}{2\(4\pi\)^2}\(\fr{T^2}{12}\)^2\bigg(\fr{3}{2}\(1+12\mubar^2\)\(1-4\mubar^2\)\(\fr{1}{\e}
+6\,\ln\,\fr{\bar{\Lambda}}{4\pi T}\)
+\fr{173}{20} + 114\mubar^2\nn
&+&132\mubar^4
- \fr{12}{5}\fr{\zeta '(-3)}{\zeta(-3)} - 96\mubar^2\fr{\zeta'(-1)}{\zeta(-1)}
- 144\Big[2\aleph(3,z) + 2\aleph(3,2z) - 4\imathb\mubar\,\aleph(2,z) \nn
&+& 8\imathb\mubar\,\aleph(2,2z)
- \(1-4\mubar^2\)\aleph(1,z)
- 8\mubar^2\aleph(1,2z) -\fr{1}{3}\imathb\mubar\(1+12\mubar^2\)\aleph(0,z)\Big] \bigg), \\
\label{m11res}
\Mt_{1,-1} &=& -\fr{3}{4\(4\pi\)^2}\(\fr{T^2}{12}\)^2\bigg(\(1+12\mubar^2\)\(1-4\mubar^2\)\(\fr{1}{\e} +
6\,\ln\,\fr{\bar{\Lambda}}{4\pi T}\)
+\fr{361}{60} + 30\mubar^2 - 100\mubar^4 \nn
&-& \fr{8}{5}\fr{\zeta'(-3)}{\zeta(-3)}-3\(1+4\mubar^2\)^2\aleph(z)
- 48\Big[8\aleph(3,z) + 12\imathb\mubar\,\aleph(2,z) \nn
&+& \(1-4\mubar^2\)\aleph(1,z)
+ \fr{1}{3}\imathb\mubar\(1-12\mubar^2\)\aleph(0,z)\Big]\bigg),\\
\N_{2,-2} &=& \!\!-8\widetilde{\cal I}_{1}^1 \widetilde{\mLARGE{\tau}}'
\!+\! \fr{1}{\(4\pi\)^2}\(\fr{T^2}{12}\)^2\bigg(\fr{4}{3}\(1+12\mubar^2\)\(1+6\mubar^2\)\(\fr{1}{\e} +
6\,\ln\,\fr{\bar{\Lambda}}{4\pi T}\)
+\fr{35}{6} + 2\gamma \nonumber
\ea
\ba
&+& 4\(29+12\gamma\)\mubar^2
+ 8\(71+36\gamma\)\mubar^4
- \fr{16}{15}\fr{\zeta '(-3)}{\zeta(-3)} - \fr{4}{3}\(1+48\mubar^2\)\fr{\zeta'(-1)}{\zeta(-1)} \nn
&-& 24\Big[8\aleph(3,z) + 6\aleph(3,2z)
- 12\imathb\mubar\,\aleph(2,z)
+ 24\imathb\mubar\,\aleph(2,2z) \nn
&-& 4\(1+2\mubar^2\)\aleph(1,z)
- 24\mubar^2\aleph(1,2z) -\imathb\mubar\(1+12\mubar^2\)\aleph(0,z)\Big]\bigg), \label{n22resa}\\
\Mt_{-2,2}
&=& -\fr{1}{\(4\pi\)^2}\(\fr{T^2}{12}\)^2\bigg(\fr{1}{12}\(29+288\mubar^2-144\mubar^4\)\(\fr{1}{\e} +
6\,\ln\,\fr{\bar{\Lambda}}{4\pi T}\)
+\fr{89}{12} + 4\gamma \nn
&+& 2\(43+24\gamma\)\mubar^2
- 68\mubar^4 + \fr{8}{3}\fr{\zeta '(-3)}{\zeta(-3)} + \fr{10}{3}\(1+\fr{84}{5}\mubar^2\)\fr{\zeta'(-1)}{\zeta(-1)} \nn
&+& 24\Big[10\,\aleph(3,z) + 18\imathb\mubar\,\aleph(2,z)
+ 2\(2-5\mubar^2\)\aleph(1,z)
+\imathb\mubar\,\aleph(0,z)\Big]\bigg), \label{m22resa} \\
I_e &=& -\fr{d_A g^4 }{4\(4\pi\)^2}\(\fr{T^2}{12}\)^2 \Bigg[2T_F \sum_{f}\bigg\{\fr{2}{3}\(5 +
72\mubar^2+144\mubar^4\)\(\fr{1}{\e}+6\,\ln\,\fr{\bar{\Lambda}}{4\pi T}\)
+ \fr{31}{3}+4\gamma \nn
&+& 8\(25+12\gamma\)\mubar^2 + 400\mubar^4
- \fr{64}{15}\fr{\zeta'(-3)}{\zeta(-3)} - \fr{16}{3}\(1+12\mubar^2\)\fr{\zeta'(-1)}{\zeta(-1)} \nn
&-& 96\Big[8\aleph(3,z) + 12\imathb\mubar\,\aleph(2,z) - \(3+8\mubar^2\)\aleph(1,z)
- \imathb\mubar\,\aleph(0,z)\Big]\bigg\} \nn
&+&\sum_{f\,g}\bigg\{576\(1+\gamma\)\mubar_f^2\mubar_g^2
-96\Big[3\(\aleph(3,z_f+z_g)+\aleph(3,z_f+z_g^*)\) \nn
&+&12\imathb\mubar_f(\aleph(2,z_f+z_g) + \aleph(2,z_f+z_g^*))
-12\mubar_g^2\,\aleph(1,z_f) -3\(\mubar_f+\mubar_g\)^2\aleph(1,z_f+z_g) \nn
&-& 3\(\mubar_f-\mubar_g\)^2\aleph(1,z_f+z_g^*)
-12\imathb\mubar_f\mubar_g^2\,\aleph(0,z_f) \Big]\bigg\}\Bigg]. \label{mufmugres}
\ea

Using the results of appendix C we furthermore get in the limit $T=0$
\ba
\widetilde{\cal I}_{1}^0 &=& -\fr{\mu^2}{8\pi^2}\(1+\e\(3+2\,\ln\,\fr{\bar{\Lambda}}{2\mu}\)\), \\
\widetilde{\cal I}_{2}^0 &=& \fr{1}{\(4\pi\)^2}\(\fr{1}{\e}+2\,\ln\,\fr{\bar{\Lambda}}{2\mu}\), \\
\widetilde{\cal I}_{2}^2 &=& \fr{\mu^2}{\(4\pi\)^2}\(1+\e\(1+2\,\ln\,\fr{\bar{\Lambda}}{2\mu}\)\), \\
\widetilde{\mLARGE{\tau}} &=& -\fr{4\mu^2}{\(4\pi\)^4}\(\fr{1}{\e}+2\(3+2\,\ln\,\fr{\bar{\Lambda}}{2\mu}\)\), \\
\N_{0,0} &=& \fr{24\mu^4}{(4\pi)^6}\(\fr{1}{\e}+6\,\ln\,\fr{\bar{\Lambda}}{2\mu}+\fr{91}{9}-\fr{16}{9}\ln\,2\), \\
\Mt_{0,0} &=& \fr{4\mu^4}{(4\pi)^6}\(\fr{1}{\e}+6\,\ln\,\fr{\bar{\Lambda}}{2\mu}+10\), \\
\N_{1,-1} &=& \fr{4\mu^4}{(4\pi)^6}\(\fr{1}{\e}+6\,\ln\,\fr{\bar{\Lambda}}{2\mu}+\fr{13}{3}+\fr{32}{3}\ln\,2\), \\
\Mt_{1,-1} &=& \fr{4\mu^4}{(4\pi)^6}\(\fr{1}{\e}+6\,\ln\,\fr{\bar{\Lambda}}{2\mu}+\fr{39}{4}\),
\ea
\ba
\Mt_{-2,2} &=& \fr{4\mu^4}{3(4\pi)^6}\(\fr{1}{\e}+6\,\ln\,\fr{\bar{\Lambda}}{2\mu}+\fr{61}{6}\).
\ea

\section{Evaluation of the hyperbolic integrals}
As a consequence of keeping $\mu$ finite, many of the one-dimensional integrals encountered in this paper differ from the
ones of \cite{az}. In addition to the ordinary hyperbolic cases we need to evaluate integrals of the type
\ba
\int_0^{\infty}{\rm d}x \, x^{z}e^{\imathb\beta x}{\rm coth}^n x\,{\rm csch}^p x,
\ea
where $n$ and $p$ are non-negative integers and $z$ and $\beta$ real numbers. This is accomplished by repeatedly applying the
relations
\ba
{\rm coth}^2x - {\rm csch}^2x = 1,
\ea
\begin{multline}
\int_0^{\infty}{\rm d}x \, x^{z}e^{\imathb\beta x}{\rm coth}^n x\,{\rm csch}^p x \;\;= \\
\fr{1}{n+p-1}\int_0^{\infty}{\rm d}x \, x^{z}e^{\imathb\beta x}{\rm coth}^{n-2}x\,{\rm csch}^p x
\bigg[n-1+\(\fr{z}{x}+\imathb\beta\){\rm tanh}\,x\bigg],
\end{multline}
and in the end performing the integrals using the results
\ba
\int_0^{\infty}{\rm d}x \, x^z e^{\imathb \beta x}{\rm coth}\,x &=& \Gamma(1+z)\(-\(-\imathb\beta\)^{-1-z} +
 2^{-z}\zeta(1+z,-\imathb\beta /2)\), \\
\int_0^{\infty}{\rm d}x \, x^z e^{\imathb \beta x}{\rm csch}\,x &=& 2^{-z}\Gamma(1+z)\zeta(1+z,1/2-\imathb\beta /2),
\ea
which can be straightforwardly derived. As in \cite{az}, UV-divergences in the individual terms of converging integrals are
regulated by introducing a factor $x^{\delta}$ in the integrand and in the end taking the limit $\delta \rightarrow 0+$.

\section{Properties of the functions $\aleph$}
In section 2 the functions $\aleph$ were defined by the formulas
\ba
\zeta'(x,y) &\equiv& \partial_x \zeta(x,y), \\
\aleph(n,w) &\equiv& \zeta'(-n,w)+\(-1\)^{n+1}\zeta'(-n,w^{*}), \\
\aleph(w) &\equiv& \Psi(w)+\Psi(w^*).
\ea
In order to analyze the behavior of the sum-integrals of appendix B at different values of $\mu$ and $T$, we need to expand these
functions in the limits of small and large $\mubar$. The results of such expansions, obtained straightforwardly
using among other things the integral representations of the zeta and digamma functions, read
\ba
\aleph(3,z) &=&
\fr{1}{480}\Big(\ln\,2-7\,\fr{\zeta'(-3)}{\zeta(-3)}\Big)+\fr{1}{24}\Big(5+6\,\ln\,2-
6\,\fr{\zeta'(-1)}{\zeta(-1)}\Big)\mubar^2 \nn &+&
\fr{1}{12}\!\(11-6\gamma-12\,\ln\,2\)\mubar^4+\mathcal{O}(\mubar^6), \\
\aleph(3,z+z') &=& \fr{1}{60}\fr{\zeta'(-3)}{\zeta(-3)}-\fr{1}{12}\Big(5-6\,\fr{\zeta'(-1)}{\zeta(-1)}\Big)
\(\mubar+\mubar'\)^2\nn
&+&
\fr{1}{12}\(11-6\gamma\)\(\mubar+\mubar'\)^4+\mathcal{O}(\mubar^6),
\ea
\ba
\aleph(2,z) &=& \fr{1}{12}\Big(1+2\,\ln\,2-2\,\fr{\zeta'(-1)}{\zeta(-1)}\Big)\imathb\mubar +
\fr{1}{3}\(3-2\gamma-4\,\ln\,2\)\imathb\mubar^3+\mathcal{O}(\mubar^5), \\
\aleph(2,z+z') &=& -\fr{1}{6}\Big(1-2\,\fr{\zeta'(-1)}{\zeta(-1)}\Big)\imathb\(\mubar+\mubar'\) +
\fr{1}{3}\(3-2\gamma\)\imathb\(\mubar+\mubar'\)^3+\mathcal{O}(\mubar^5), \nn
\aleph(1,z) &=& -\fr{1}{12}\Big(\ln\,2-\fr{\zeta'(-1)}{\zeta(-1)}\Big) - \(1-2\,\ln\,2-\gamma\)\mubar^2
-\fr{7}{6}\zeta(3)\,\mubar^4+\mathcal{O}(\mubar^6), \\
\aleph(1,z+z') &=& -\fr{1}{6}\fr{\zeta'(-1)}{\zeta(-1)}-\(1-\gamma\)\(\mubar+\mubar'\)^2-
\fr{1}{6}\zeta(3)\(\mubar+\mubar'\)^4+\mathcal{O}(\mubar^6), \\
\aleph(0,z) &=& 2\(2\,\ln\,2+\gamma\)\imathb\mubar-\fr{14}{3}\zeta(3)\,\imathb\mubar^3+\mathcal{O}(\mubar^5), \\
\aleph(z) &=& -2\gamma -4\,\ln\,2 + 14\,\zeta(3)\,\mubar^2 - 62\,\zeta(5)\,\mubar^4+\mathcal{O}(\mubar^5)
\ea
for small $\mubar$ and
\ba
\aleph(3,z) &=& \fr{1}{2}\mubar^4\Big(\ln\,\mubar-\fr{1}{4}\Big)+\fr{1}{4}\mubar^2\Big(\ln\,\mubar+\fr{1}{3}\Big) +
\mathcal{O}(\ln\,\mubar), \\
\aleph(2,z) &=& \fr{2i}{3}\mubar^3\Big(\ln\,\mubar-\fr{1}{3}\Big) +\fr{i}{6}\mubar\Big(\ln\,\mubar+\fr{1}{2}\Big)
+\mathcal{O}(\fr{\ln\,\mubar}{\mubar}),\\
\aleph(1,z) &=& -\mubar^2\Big(\ln\,\mubar-\fr{1}{2}\Big)-\fr{1}{12}\Big(\ln\,\mubar+1\Big) +
\mathcal{O}(\fr{\ln\,\mubar}{\mubar^2}),\\
\aleph(0,z) &=& -2i\mubar\(\ln\,\mubar-1\)-\fr{i}{12}\fr{1}{\mubar}+\mathcal{O}(\fr{\ln\,\mubar}{\mubar^3}), \\
\aleph(z) &=& 2\,\ln\,\mubar -\fr{1}{12}\fr{1}{\mubar^2}+\mathcal{O}(\fr{\ln\,\mubar}{\mubar^4})
\ea
for large $\mubar$. 

\section{The diagram $I_e'$ and the plasmon term at $T=0$}
In section 4, where the zero-temperature pressure of QCD was computed to $\mathcal{O}(g^4)$, we only quoted the results for the diagram $I_e'$ and the plasmon term. The corresponding calculations will be performed in this appendix.

\subsection{The diagram $I_e'$}
The one-loop gluon polarization tensor depicted in Fig. 4.b. has in the Feynman gauge the expression
\ba
\(\Pi_{\mu\nu}^{\rmi{f}}\)^{ab}(P)&=& -2g^2T_F \delta^{ab}\bigg(2\It_1\delta_{\mu\nu}+\(P_{\mu}P_{\nu}-P^2\delta_{\mu\nu}\)\Pi_{\rm f} \!\(P\)-
\sumint_{\{Q\}}\fr{(2Q-P)_{\mu}(2Q-P)_{\nu}}{Q^2(Q-P)^2}\bigg). \label{polar}
\ea
Denoting its vacuum ($T=\mu=0$) part here by $\(\Pi_{\mu\nu}^{\rmi{f}}\)^{ab}(P)\mid_{\rmi{vac}}$ we obtain after a
straightforward computation
\ba
\(\Pi_{\mu\nu}^{\rmi{f}}\)^{ab}(P)\mid_{\rmi{vac}}&=&-2g^2AT_f\delta^{ab}\(\fr{\Lambda^2}{P^2}\)^{\!\!\e}\(P_{\mu}P_{\nu}-
P^2\delta_{\mu\nu}\), \label{polarvac}
\ea
where the coefficient $A$ can be shown to have the the $\e$ expansion
\ba
A &=& \fr{1}{24\pi^2}\(\fr{1}{\e}-\gamma+\ln (4\pi)+\fr{5}{3}+\mathcal{O}(\e)\).
\ea
Using this expression we easily obtain at $T=0$
\ba
I_e' &=&
4(d-2)g^4Ad_AT_F^2\Lambda^{2\e}\int_{P\{Q\}}\fr{1}{(P^2)^{\e}Q^2(P-Q)^2}\;\,\,\equiv\;\;\, 4(d-2)g^4Ad_AT_F^2\Lambda^{2\e}\widetilde{\mLARGE{\tau}}'', \label{dia01}
\ea
and only the integral $\widetilde{\mLARGE{\tau}}''$ remains to be evaluated.

The most straightforward way to tackle the computation of the new integral is to proceed as in the case of the sum-integral $\widetilde{\mLARGE{\tau}}$, while from the beginning on neglecting terms that vanish at $T=0$. Using Eq. (\ref{piflp}) we get
\ba
\widetilde{\mLARGE{\tau}}''|_{T=0} &=&\int_P \fr{1}{(P^2)^{\e}}\Pi_{{\rm f},T=0}\!\(P\)\;\;\,=\;\;\,
\Lambda^{-2\e}\Bigg\{\int_P \(\(\fr{\Lambda^2}{P^2}\!\!\)^{\e}-1\)\Pi_{{\rm f},T=0}\!\(P\)+\int_P \Pi_{{\rm f},T=0}\!\(P\)\Bigg\}\nn
&=& \Lambda^{-2\e}\Bigg\{\(\widetilde{\cal I}_{1}^0\)^2+\beta_2 T^4\int_P \(\(\fr{\Lambda^2}{P^2}\!\!\)^{\e}-1\)\(\fr{1}{P^4}-\(4-2\e\)\fr{p_0^2}{P^2}\)\nn
&-&\fr{\e \,T}{\(4\pi\)^2}\int_P\ln\,\fr{P^2}{\Lambda^2}\!\int\! {\rm d}^3 r\,\fr{e^{\imathb \mathbf{p} \cdot \mathbf{r}}}{r^2}\!
\bigg[\cos(2\bar{\mu} \bar{r}) {\rm csch} \,\bar{r} -\fr{1}{\bar{r}}+\(\fr{1}{6}+2\bar{\mu}^2\)\bar{r}\nn
&-&\(\fr{7}{360}+\fr{\bar{\mu}^2}{3}+\fr{2\bar{\mu}^4}{3}\)\bar{r}^3\bigg]e^{-|p_0\!| r}+\mathcal{O}(\e)\Bigg\}{\mbox{\Huge{$\mid$}}}_{T=0},
\ea
where the remaining $P$-integrals can be evaluated using previous results. The calculation of the $r$-integral then leads after
setting the temperature to zero to
\ba
\widetilde{\mLARGE{\tau}}'' &=&\fr{4\mu^{4-2\e}}{(4\pi)^4}
\(1-2\e\Big[3\,\ln\,2-4-2\,\ln\,\fr{\bar{\Lambda}}{\mu}\Big]+\mathcal{O}(\e^2)\), \label{taupp}
\ea
and plugging this expression to Eq. (\ref{dia01}) gives Eq. (\ref{pqcd2res}) as the result.

\subsection{The plasmon term}
The evaluation of the plasmon contribution to the $T=0$ pressure corresponds to summing over all the ring diagrams of
Fig. 4.c. starting at three-loop order. These graphs contain as loop insertions the fermionic part of the vacuum
subtracted one-loop gluon polarization tensor, which we denote by $\(\Delta \Pi_{\mu\nu}^{\rmi{f}}\)^{ab}$. The summation was
originally performed in \cite{fmcl} and will be reproduced here following to a large extent the treatment of \cite{fmcl,tt2}. The
computation begins from the derivation of an integral representation for the polarization tensor and then proceeds
to the evaluation of the actual plasmon sum.

Using Eqs. (\ref{polar}) and (\ref{polarvac}) it is easily seen that the vacuum subtracted polarization tensor is orthogonal to the external momentum, i.e.
\ba
P_{\mu}\(\Delta \Pi_{\mu\nu}^{\rmi{f}}\)^{ab}(P) &\equiv&P_{\mu}\Big(\(\Pi_{\mu\nu}^{\rmi{f}}\)^{ab}(P)-\(\Pi_{\mu\nu}^{\rmi{f}}\)^{ab}(P)\mid_{\rmi{vac}}\Big)\;\;\,=\;\;\, 0.
\ea
Setting $\e=0$ and applying this relation together with rotational invariance we may then write the tensor in the form
\ba
\(\Delta\Pi_{\mu\nu}^{\rmi{f}}\)^{ab}(P)&=&\fr{1}{p^2}\(\Delta\Pi_{00}^{\rmi{f}}\)^{ab}(P)(P^2\delta_{\mu\nu}-P_{\mu}P_{\nu}) \nn
&+&\fr{1}{2p^2}\(\(\Delta\Pi_{\mu\mu}^{\rmi{f}}\)^{ab}(P)-\fr{3P^2}{p^2}\(\Delta\Pi_{00}^{\rmi{f}}\)^{ab}(P)\)\delta_{\mu i}\delta_{\nu j} (p^2\delta_{ij}-p_i p_j) \\
&=&\(\Delta\Pi_{00}^{\rmi{f}}\)^{ab}(P)\fr{P^2}{p^2}\bigg[\delta_{\mu\nu}-\fr{P_{\mu}P_{\nu}}{P^2} - \delta_{\mu i}\delta_{\nu j} \(\delta_{ij}-\fr{p_i p_j}{p^2}\)\bigg] \nn
&+& \fr{1}{2}\(\(\Delta\Pi_{\mu\mu}^{\rmi{f}}\)^{ab}(P) - \fr{P^2}{p^2}\(\Delta\Pi_{00}^{\rmi{f}}\)^{ab}(P)\)\delta_{\mu i}\delta_{\nu j} \(\delta_{ij}-\fr{p_i p_j}{p^2}\), \label{deltapi}
\ea
where $P$ stands for four-vectors and $p$ for three-vectors as before and the polarization tensor has been divided into two parts proportional to orthonormal projection operators. The components of the tensor appearing explicitly in the result are available through an application of the Residue theorem, which after a straightforward calculation leads to the results (see appendix A of part II in \cite{fmcl})
\ba
\(\Delta\Pi_{00}^{\rmi{f}}\)^{ab}(P) &=& 4g^2\delta^{ab}\sum_f \int\!\fr{{\rm d}^3q}{(2\pi)^3}\fr{1}{q} \theta(\mu-q)\fr{p^2q^2-(p\cdot q)^2}{(Q-P)^2(Q+P)^2}{\mbox{\huge{$\mid$}}}_{q_0=iq}, \label{pi00x} \\
\(\Delta\Pi_{\mu\mu}^{\rmi{f}}\)^{ab}(P) &=& 4g^2\delta^{ab}\sum_f\int\!\fr{{\rm d}^3q}{(2\pi)^3}\fr{1}{q}\theta(\mu-q) \fr{(P\cdot Q)^2}{(Q-P)^2(Q+P)^2}{\mbox{\huge{$\mid$}}}_{q_0=iq}. \label{pimumux}
\ea

Defining an angular variable $\phi$ by
\ba
\fr{p_0}{\mid\! {\mbox{\boldmath$p$}}\! \mid }&=&{\rm arc\,tan}\,\phi
\ea
and performing the corresponding integrations in Eqs. (\ref{pi00x}) and (\ref{pimumux}), we now obtain for the polarization tensor
\ba
\(\Delta\Pi_{00}^{\rmi{f}}\)^{ab}(P) &=& \fr{g^2\delta^{ab}}{2\pi^2} \int_0^{\mu}\!\!\!{\rm d}q
q\bigg\{1+\fr{4q^2-P^2}{8qP\sin\phi}\ln\fr{P^2+4q^2+4qP\sin\phi}{P^2+4q^2-4qP\sin\phi} -
\cot\phi\, {\rm arc\,tan} \fr{4q^2\sin2\phi}{P^2+4q^2\cos2\phi} \bigg\} \nn
&=& \fr{g^2\delta^{ab}}{2\pi^2}\bigg\{\fr{2}{3}\mu^2 +
\fr{\mu(4\mu^2-3P^2)}{24P\sin\phi}\ln\fr{P^2+4\mu^2+4\mu P\sin\phi}{P^2+4\mu^2-4\mu P\sin\phi}\nn
&-&\fr{P^2\sin^2\phi}{24}\ln\bigg[1+8\,\fr{\mu^2P^2\cos2\phi+2\mu^4}{P^4}\bigg]\nn
&-&\fr{1}{2}\(\mu^2-\fr{1+2\sin^2\phi}{12}P^2\)\cot\phi\, {\rm arc\,tan}
\fr{\sin 2\phi}{\cos 2\phi+P^2/(4\mu^2)} \bigg\}, \label{pi00y} \\
\(\Delta\Pi_{\mu\mu}^{\rmi{f}}\)^{ab}(P) &=& \fr{g^2\delta^{ab}}{\pi^2} \int_0^{\mu}\!\!\!{\rm d}q
q\bigg\{1-\fr{P}{8q\sin\phi}\ln\fr{P^2+4q^2+4qP\sin\phi}{P^2+4q^2-4qP\sin\phi} \bigg\} \nn
&=&  \fr{g^2\delta^{ab}}{2\pi^2}\bigg\{\mu^2 -
\fr{\mu P}{4\sin\phi}\ln\fr{P^2+4\mu^2+4\mu P\sin\phi}{P^2+4\mu^2-4\mu P\sin\phi}
-\fr{P^2}{8}\ln\bigg[1+8\,\fr{\mu^2P^2\cos2\phi+2\mu^4}{P^4}\bigg]\nn
&+&\fr{P^2\cot\phi}{4}\, {\rm arc\,tan}
\fr{\sin 2\phi}{\cos 2\phi+P^2/(4\mu^2)} \bigg\},\label{pimumuy}
\ea
where $P$ this time stands for the norm of the corresponding four-vector and where flavor sums have been suppressed. The last
forms obtained here are, however, inconvenient to work with. It is on the other hand easy to confirm by a direct
integration that the simple integral representations
\ba
\(\Delta\Pi_{00}^{\rmi{f}}\)^{ab}(P) &=& \fr{g^2\delta^{ab}}{2\pi^2}\fr{\mu^4\sin^2\phi}{P^2}\int_0^1\!\!
{\rm d}x\int_{-1}^1 \!\!\!{\rm d}y\fr{x(1-y^2)}{1-4x\mu^2/P^2\sin^2\phi(y+i\cot\phi)^2}, \label{pi00z}\\
\(\Delta\Pi_{\mu\mu}^{\rmi{f}}\)^{ab}(P) &=& -\fr{g^2\delta^{ab}}{\pi^2}\fr{\mu^4\sin^2\phi}{P^2}\int_0^1\!\!
{\rm d}x\int_{-1}^1\!\!\! {\rm d}y\fr{x(y+i\cot\phi)^2}{1-4x\mu^2/P^2\sin^2\phi(y+i\cot\phi)^2} \label{pimumuz}
\ea
produce exactly Eqs. (\ref{pi00y}) and (\ref{pimumuy}), which suggests the use of the Sommerfeld-Watson integral
formula
\ba
\fr{1}{1+x}\;\;\,=\;\;\,-\int_{-i\infty-\e}^{i\infty-\e} \fr{{\rm d}z}{2\pi i} \fr{\pi}{\sin\pi z}x^z.
\ea

When applied to Eqs. (\ref{pi00z}) and (\ref{pimumuz}), the Sommerfeld-Watson formula gives
\ba
\(\Delta\Pi_{00}^{\rmi{f}}\)^{ab}(P) &=& \fr{g^2\mu^2\delta^{ab}}{8\pi^2}\int_{-i\infty-\e}^{i\infty-\e}
\fr{{\rm d}z}{2\pi i}\fr{\pi}{\sin\pi z}\nn
&\times&\int_0^1\!\!{\rm d}x\int_{-1}^1 \!\!\!{\rm d}y\fr{(1-y^2)}{(y+i\cot\phi)^2}
\bigg[\fr{-P^2}{4x\mu^2\sin^2\phi(y+i\cot\phi)^2}\bigg]^z, \label{pi00a} \\
\(\Delta\Pi_{\mu\mu}^{\rmi{f}}\)^{ab}(P) &=& -\fr{g^2\mu^2\delta^{ab}}{4\pi^2}\int_{-i\infty-\e}^{i\infty-\e}
\fr{{\rm d}z}{2\pi i}\fr{\pi}{\sin\pi z}\int_0^1\!\!{\rm d}x\int_{-1}^1 \!\!\!{\rm d}y\bigg[\fr{-P^2}{4x\mu^2\sin^2\phi(y+i\cot\phi)^2}\bigg]^z, \label{pimumua}
\ea
where the $x$ and $y$ integrals can now be factorized and performed separately. The result first derived in \cite{fmcl} is that
one has obtained a compact integral representation for the polarization tensor
\ba
\(\Delta\Pi_{00}^{\rmi{f}}\)^{ab}(P) &=&-\fr{g^2\delta^{ab}}{2\pi^2}\sum_f \,\mu^2\!\int_{-i\infty-\e}^{i\infty-\e}
\fr{{\rm d}z}{2\pi i} \Gamma_1(z,\phi)\(\fr{P^2}{4\mu^2}\)^z, \label{pi00c}\\
\(\Delta\Pi_{\mu\mu}^{\rmi{f}}\)^{ab}(P) &=& -\fr{g^2\delta^{ab}}{2\pi^2}\sum_f \,\mu^2\!\int_{-i\infty-\e}^{i\infty-\e}
\fr{{\rm d}z}{2\pi i} \Gamma_2(z,\phi)\(\fr{P^2}{4\mu^2}\)^z, \label{pimumuc}
\ea
where flavor-sums have been re-introduced and the functions $\Gamma_n$ are defined by
\ba
\Gamma_1(z,\phi)&=&\fr{\pi}{\sin\pi z}\fr{\cos (2z\phi)-1/(2z)\sin (2z\phi) \cot\phi}{(1-z)(1-4z^2)}, \\
\Gamma_2(z,\phi)&=&\fr{\pi}{\sin\pi z}\fr{\cos (2z\phi)-\sin (2z\phi) \cot\phi}{(1-z)(1-2z)}.
\ea

We are now ready to evaluate the sum of the diagrams of Fig. 4.c, which we begin by first defining (again in analogy with \cite{fmcl})
\ba
\Lambda_1(P)&=&\fr{1}{d_A\sin^2\phi}\(\Delta\Pi_{00}^{\rmi{f}}\)^{aa}(P),\\
\Lambda_2(P)&=&\fr{1}{2d_A}\(\(\Delta\Pi_{\mu\mu}^{\rmi{f}}\)^{aa}(P) -\fr{1}{\sin^2\phi} \(\Delta\Pi_{00}^{\rmi{f}}\)^{aa}(P)\).
\ea
Using the orthonormality of the projection operators appearing in Eq. (\ref{deltapi}) and keeping in mind the symmetry factors for the
ring diagrams, this straightforwardly leads to the expression
\ba
p_\rmi{3}&=&-\fr{d_A}{2}\int \fr{{\rm d}^4 P}{(2\pi)^4}\bigg\{\ln \bigg[1+\fr{\Lambda_1(P)}{P^2}\bigg]-\fr{\Lambda_1(P)}{P^2} + 2\,\ln\bigg[1+\fr{\Lambda_2(P)}{P^2}\bigg]-2\fr{\Lambda_2(P)}{P^2}\bigg\}, \label{andy1}
\ea
which we now must examine.

The integrals over the three-dimensional angular variables are trivially performed. This gives for the infrared sensitive part of the plasmon contribution
\ba
p_\rmi{3}^a &\equiv& -\fr{d_A}{2}\int \fr{{\rm d}^4 P}{(2\pi)^4}\bigg\{\ln \bigg[1+\fr{\Lambda_1(0,\phi)}{P^2}\bigg] + 2\,\ln\bigg[1+\fr{\Lambda_2(0,\phi)}{P^2}\bigg]-\fr{1}{P^2}\Big(\Lambda_1(0,\phi)+2\Lambda_2(0,\phi)\Big)\nn
&+&\fr{1}{2P^2}\fr{1}{P^2+4\boldmu^2}\Big(\Lambda_1^2(0,\phi)+\Lambda_2^2(0,\phi)\Big)\bigg\} \\
&=& -\fr{d_A}{(2\pi)^3}\int_0^{\infty} \!\!\!\!{\rm d} P^2 P^2\int_0^{\pi/2} \!\!\!\!\!\!{\rm d}\phi \sin^2 \phi \bigg\{\ln \bigg[1+\fr{\Lambda_1(0,\phi)}{P^2}\bigg] + 2\,\ln\bigg[1+\fr{\Lambda_2(0,\phi)}{P^2}\bigg] \nn
&-& \fr{1}{P^2}\Big(\Lambda_1(0,\phi)+2\Lambda_2(0,\phi)\Big) + \fr{1}{2P^2}\fr{1}{P^2+4\boldmu^2}\Big(\Lambda_1^2(0,\phi)+\Lambda_2^2(0,\phi)\Big)\bigg\} \\
&=& -\fr{d_A}{2(2\pi)^3}\int_0^{\pi/2} \!\!\!\!\!\!{\rm d}\phi \sin^2 \phi\bigg\{\Lambda_1^2(0,\phi)\(\ln\fr{\Lambda_1(0,\phi)}{4\boldmu^2}-\fr{1}{2}\) + 2\Lambda_2^2(0,\phi)\(\ln\fr{\Lambda_2(0,\phi)}{4\boldmu^2}-\fr{1}{2}\)\bigg\},
\ea
where the $P^2$-integral has been evaluated in the last stage. From Eqs. (\ref{pi00y}) and (\ref{pimumuy}) we furthermore have
\ba
\Lambda_1(0,\phi) &=& \fr{g^2\boldmu^2}{2\pi^2}\fr{1-\phi\cot\phi}{\sin^2\phi}, \\
\Lambda_2(0,\phi) &=& \fr{g^2\boldmu^2}{4\pi^2}\(1-\fr{1-\phi\cot\phi}{\sin^2\phi}\),
\ea
which gives
\ba
p_\rmi{3}^a &=& -\fr{2d_A(\boldmu^2)^2}{\pi^3}\bigg(\!\fr{g}{4\pi}\!\!\!\bigg)^{\!\!4}
\int_0^{\pi/2} \!\!\!\!\!\!{\rm d}\phi \sin^2 \phi
\bigg\{2\(\fr{1-\phi\cot\phi}{\sin^2\phi}\)^2\(\ln\bigg[\fr{g^2}{8\pi^2}\fr{1-\phi\cot\phi}{\sin^2\phi}\bigg]
-\fr{1}{2}\) \nn
&+& \(1-\fr{1-\phi\cot\phi}{\sin^2\phi}\)^2\(\ln\bigg[\fr{g^2}{16\pi^2}\(1-\fr{1-\phi\cot\phi}{\sin^2\phi}\)\bigg]-\fr{1}{2}\)\bigg\}.
\ea
Performing the remaining $\phi$-integral produces now
\ba
p_\rmi{3}^a &=& -\fr{d_A(\boldmu^2)^2}{4\pi^2}\(2\,\ln\fr{g^2}{16\pi^2} - 1
+ \fr{16}{3}\ln\,2\(1-\ln\,2\)+\delta\) \label{ppl1res}
\ea
with the constant $\delta$ defined in Eq. (\ref{deltadef}) of section 4.

The second part of the plasmon term is defined by
\ba
p_\rmi{3}^b&\equiv&p_\rmi{3}-p_\rmi{3}^a.
\ea
To order $g^4$ it can be evaluated by simply expanding the logarithms in Eq. (\ref{andy1}) in powers of $g^2$, since there obviously will be
no $g^4\ln\,g$ contributions originating from its expression
\ba
p_\rmi{3}^b &\equiv& -\fr{d_A}{(2\pi)^3}\int_0^{\infty} \!\!\!\!{\rm d} P^2 P^2\int_0^{\pi/2} \!\!\!\!\!\!{\rm d}\phi
\sin^2 \phi \bigg \{ \ln\bigg[\fr{P^2+\Lambda_1(P)}{P^2+\Lambda_1(0,\phi)}\bigg] +
2\,\ln\bigg[\fr{P^2+\Lambda_2(P)}{P^2+\Lambda_2(0,\phi)}\bigg] \nn
&-& \fr{1}{P^2}\Big(\Lambda_1(P)+2\Lambda_2(P)-\Lambda_1(0,\phi)-2\Lambda_2(0,\phi)\Big) -
\fr{1}{2P^2}\fr{1}{P^2+4\boldmu^2}\Big(\Lambda_1^2(0,\phi)+\Lambda_2^2(0,\phi)\Big)\bigg\} \nn
&=& \fr{d_A}{2(2\pi)^3}\int_0^{\infty} \!\!\!\!{\rm d} P^2 \int_0^{\pi/2} \!\!\!\!\!\!{\rm d}\phi \sin^2 \phi \bigg \{
\fr{1}{P^2}\Big(\Lambda_1^2(P)+2\Lambda_2^2(P)\Big) \nn
&+& \(\fr{1}{P^2+4\boldmu^2}-\fr{1}{P^2}\)\Big(\Lambda_1^2(0,\phi)+2\Lambda_2^2(0,\phi)\Big)\bigg\} +\mathcal{O}(g^5).
\ea
Plugging in the contour-integral representations of the $\Lambda$-functions,
\ba
\Lambda_1(P) &=& -\fr{g^2}{2\pi^2\sin^2\phi}\sum_f \mu_f^2\int_{-i\infty-\e}^{i\infty-\e}
\fr{{\rm d}z}{2\pi i} \Gamma_1(z,\phi)\(\fr{P^2}{4\mu_f^2}\)^z\nn
&\equiv&\sum_f\int_{-i\infty-\e}^{i\infty-\e}
\fr{{\rm d}z}{2\pi i} \widetilde{\Lambda}_1^f(z,\phi)\(\fr{P^2}{4\mu_f^2}\)^z\!\!, \label{la1til} \\
\Lambda_2(P) &=& -\fr{g^2}{4\pi^2}\sum_f \mu_f^2\int_{-i\infty-\e}^{i\infty-\e}
\fr{{\rm d}z}{2\pi i} \bigg\{\Gamma_2(z,\phi)-\fr{\Gamma_1(z,\phi)}{\sin^2\phi}\bigg\}\(\fr{P^2}{4\mu_f^2}\)^z \nn
&\equiv& \sum_f\int_{-i\infty-\e}^{i\infty-\e}
\fr{{\rm d}z}{2\pi i} \widetilde{\Lambda}_2^f(z,\phi)\(\fr{P^2}{4\mu_f^2}\)^z, \label{la2til}
\ea
and performing the $P^2$-integral with an IR cutoff $4\eta\boldmu^2$, we now obtain
\ba
p_\rmi{3}^b &=& \fr{d_A}{2(2\pi)^3}\sum_{fg}\int_{4\eta{\mbox{\small{\boldmath$\mu$}}}^2}^{\infty} \!\!\!\!{\rm d} P^2 \int_{0}^{\pi/2} \!\!\!\!\!\!{\rm d}\phi \sin^2 \phi \Bigg\{ \fr{1}{n_f^2}\(\fr{1}{P^2+4\boldmu^2}-\fr{1}{P^2}\)\Big(\Lambda_1^2(0,\phi)+2\Lambda_2^2(0,\phi)\Big) \\
&+&\fr{1}{P^2} \int_{-i\infty-\e}^{i\infty-\e}\fr{{\rm d}z}{2\pi i} \int_{-i\infty-\e'}^{i\infty-\e'}
\fr{{\rm d}z'}{2\pi i} \Big(\widetilde{\Lambda}_1^f(z,\phi)\widetilde{\Lambda}_1^g(z',\phi) +
2\widetilde{\Lambda}_2^f(z,\phi)\widetilde{\Lambda}_2^g(z',\phi)\Big)\(\fr{P^2}{4\mu_f^2}\)^{z}
\Bigg(\fr{P^2}{4\mu_g^2}\Bigg)^{z'} \Bigg\} \nn
&\equiv& \fr{d_A}{2(2\pi)^3}\sum_{fg} \int_{0}^{\pi/2} \!\!\!\!\!\!{\rm d}\phi \sin^2 \phi
\Bigg\{ \fr{1}{n_f^2}\Big(\Lambda_1^2(0,\phi)+2\Lambda_2^2(0,\phi)\Big)\ln\,\eta \\
&-& \int_{-i\infty-\e}^{i\infty-\e}\fr{{\rm d}z}{2\pi i} \int_{-i\infty-\e'}^{i\infty-\e'}\fr{{\rm d}z'}{2\pi i}
\fr{\eta^{z+z'}}{z+z'}\Big(\widetilde{\Lambda}_1^f(z,\phi)\widetilde{\Lambda}_1^g(z',\phi) +
2\widetilde{\Lambda}_2^f(z,\phi)\widetilde{\Lambda}_2^g(z',\phi)\Big)\(\fr{\boldmu^2}{\mu_f^2}\)^{z}
\Bigg(\fr{\boldmu^2}{\mu_g^2}\Bigg)^{z'} \Bigg\}. \nonumber
\ea

The $\phi$-integral in the first term is easy to perform, but is left in its present form for now. Instead, we deform the integration contour of the $z'$ integral in order to be able to proceed to the limit $\eta\rightarrow 0$. Denoting Residues by $\Re$ and encircling the poles at $z'=0$ and $z'=-z$ to obtain a contour-integral, which vanishes in this limit, we get
\ba
&&\int_{-i\infty-\e'}^{i\infty-\e'}\fr{{\rm d}z'}{2\pi i}
\fr{\eta^{z+z'}}{z+z'}\Big(\widetilde{\Lambda}_1^f(z,\phi)\widetilde{\Lambda}_1^g(z',\phi) +
2\widetilde{\Lambda}_2^f(z,\phi)\widetilde{\Lambda}_2^g(z',\phi)\Big)\(\fr{\boldmu^2}{\mu_f^2}\)^{z}\Bigg(\fr{\boldmu^2}{\mu_g^2}\Bigg)^{z'} \nn
&=&-\fr{\eta^{z}}{z}\Big(\widetilde{\Lambda}_1^f(z,\phi)\Re\widetilde{\Lambda}_1^g(z',\phi)\mid_{z'=0} +
2\widetilde{\Lambda}_2^f(z,\phi)\Re\widetilde{\Lambda}_2^g(z',\phi)\mid_{z'=0}\Big)\(\fr{\boldmu^2}{\mu_f^2}\)^{z}\nn
&-&\Big(\widetilde{\Lambda}_1^f(z,\phi)\widetilde{\Lambda}_1^g(-z,\phi) +
2\widetilde{\Lambda}_2^f(z,\phi)\widetilde{\Lambda}_2^g(-z,\phi)\Big)\Bigg(\fr{\mu_g^2}{\mu_f^2}\Bigg)^{z}+\mathcal{O}(\eta^{\e}). \label{pole1}
\ea
The $z$-integral of the first term is quickly evaluated by closing the contour on the right half-plane, where the only pole is located at the origin $z=0$. The substitution $\eta^{z}=1+z\,\ln\,\eta+\mathcal{O}(z^2)$ then produces
\ba
p_\rmi{3}^b &=& \fr{d_A}{2(2\pi)^3}\sum_{fg} \int_{0}^{\pi/2} \!\!\!\!\!\!{\rm d}\phi \sin^2 \phi
\bigg\{ \int_{-i\infty-\e}^{i\infty-\e}\fr{{\rm d}z}{2\pi i}\Big(\widetilde{\Lambda}_1^f(z,\phi)\widetilde{\Lambda}_1^g(-z,\phi) +
2\widetilde{\Lambda}_2^f(z,\phi)\widetilde{\Lambda}_2^g(-z,\phi)\Big)\(\fr{\mu_g^2}{\mu_f^2}\)^{z} \nn
&-&\Re^2\bigg[\fr{1}{z}\Big(\widetilde{\Lambda}_1^f(z,\phi)\widetilde{\Lambda}_1^g(z',\phi) +
2\widetilde{\Lambda}_2^f(z,\phi)\widetilde{\Lambda}_2^g(z',\phi)\Big)
\(\fr{\boldmu^2}{\mu_f^2}\)^{z}\bigg]\mid_{z=z'=0}\bigg\}\;\;\equiv\;\;p_\rmi{3}^{b,1}+p_\rmi{3}^{b,2}, \label{pplas2}
\ea
where we have used the relation
\ba
\sum_{fg}\Re^2\Big\{\widetilde{\Lambda}_1^f(z,\phi)\widetilde{\Lambda}_1^g(z',\phi) +
2\widetilde{\Lambda}_2^f(z,\phi)\widetilde{\Lambda}_2^g(z',\phi)\Big\}\mid_{z=z'=0}
&=& \Lambda_1^2(0,\phi)+2\Lambda_2^2(0,\phi),
\ea
a trivial consequence of the definitions of Eqs. (\ref{la1til}) and (\ref{la2til}).

The evaluation of the plasmon term has now been reduced to the computation of the two integrals $p_\rmi{3}^{b,1}$ and $p_\rmi{3}^{b,2}$. Using the relations
\ba
\widetilde{\Lambda}_1(z,\phi)&=&-\fr{g^2\mu^2}{2\pi^2\sin^2\phi}\fr{\pi}{\sin\,\pi z}\fr{\cos\,2z\phi-\sin\,2z\phi\cot\phi /2z}{(1-z)(1-4z^2)}, \\
\widetilde{\Lambda}_2(z,\phi)&=&-\fr{g^2\mu^2}{4\pi^2}\fr{\pi}{\sin\,\pi z}\(\fr{\cos\,2z\phi-\sin\,2z\phi\cot\phi }
{(1-z)(1-2z)} - \fr{\cos\,2z\phi-\sin\,2z\phi\cot\phi /2z}{(1-z)(1-4z^2)\sin^2\phi}\),
\ea
\ba
\Re\widetilde{\Lambda}_1(z,\phi)\mid_{z=0}&=&-\fr{g^2\mu^2}{2\pi^2\sin^2\phi}\(1-\phi\cot\phi\), \\
\Re\widetilde{\Lambda}_2(z,\phi)\mid_{z=0}&=&-\fr{g^2\mu^2}{4\pi^2\sin^2\phi}\(\sin^2\phi-1+\phi\cot\phi\)
\ea
the double residue appearing in $p_\rmi{3}^{b,2}$ is rapidly evaluated, and the computation of the subsequent $\phi$-integral produces
\ba
p_\rmi{3}^{b,2}&=&-\fr{d_A\boldmu^2}{\pi^2}\bigg(\!\fr{g}{4\pi}\!\!\!\bigg)^{\!\!4}\bigg\{\(\fr{\pi^2}{6}-1\)\boldmu^2 -
\fr{1}{2}\sum_f \mu^2\,\ln\,\fr{\mu^2}{\boldmu^2}\bigg\}. \label{ppl2bres}
\ea

For $p_\rmi{3}^{b,1}$ the $\phi$-integration can, on the other hand, be immediately performed. This gives
\ba
p_\rmi{3}^{b,1} &=& -\fr{d_A}{2\pi^2}\bigg(\!\fr{g}{4\pi}\!\!\!\bigg)^{\!\!4}\sum_{fg}\mu_f^2\mu_g^2\int_{-i\infty-\e}^{i\infty-\e} \fr{{\rm d}z}{2\pi i}\fr{\pi\cot\pi z}{z(1-z^2)(1-4z^2)} \Bigg(\fr{\mu_g^2}{\mu_f^2}\Bigg)^{z}, \label{p3b1}
\ea
where the remaining contour integral is straightforward to evaluate by applying the Residue theorem. Assuming first $\mu_f^2\geq\mu_g^2$ we may close the integration contour by a semi-circle on the right half-plane, where the integrand obviously has poles located at $z=k,\,\forall k\in {\rm N}$. Denoting $\mu_g^2/\mu_f^2\equiv\beta$ we get
\ba
&&\int_{-i\infty-\e}^{i\infty-\e} \fr{{\rm d}z}{2\pi i}\fr{\beta^z\pi\cot\pi z}{z(1-z^2)(1-4z^2)} \nn
&=&-\ln\,\beta+\fr{\beta}{6}\(\fr{25}{6}-\ln\,\beta\) - \sum_{k=2}^{\infty}\fr{\beta^k}{k(k^2-1)(4k^2-1)} \nn
&=&-\ln\,\beta+\fr{\beta}{6}\(\fr{25}{6}-\ln\,\beta\) -\(\fr{7}{6}+\fr{25\beta}{36}-\fr{1+6\beta+\beta^2}{6\beta}\ln [1-\beta] - \fr{2(1+\beta)}{3\sqrt{\beta}}\ln\fr{1+\sqrt{\beta}}{1-\sqrt{\beta}}\) \nn
&=&-\fr{7}{6}-\fr{6+\beta}{6}\ln\,\beta + \fr{4(1+\beta)}{3\sqrt{\beta}}\ln [1+\sqrt{\beta}]+\fr{(1-\sqrt{\beta})^4}{6\beta}\ln [1-\beta], \label{p2aint1}
\ea
where the infinite sum has been evaluated using standard formulae. For $\mu_f^2\leq\mu_g^2 \,\Leftrightarrow\,\beta\geq 1$ the computation proceeds in an analogous fashion the only difference being that the integration contour is now closed on the left half-plane. The result of this calculation is simply
\ba
\int_{-i\infty-\e}^{i\infty-\e} \fr{{\rm d}z}{2\pi i}\fr{\beta^z\pi\cot\pi z}{z(1-z^2)(1-4z^2)}
=-\fr{7}{6}-\fr{6+\beta}{6}\ln\,\beta + \fr{4(1+\beta)}{3\sqrt{\beta}}\ln [1+\sqrt{\beta}]+\fr{(1-\sqrt{\beta})^4}{6\beta}\ln [\beta-1], \label{p2aint2}
\ea
and the application of Eqs. (\ref{p2aint1}) and (\ref{p2aint2}) in Eq. (\ref{p3b1}) gives now
\ba
p_\rmi{3}^{b,1} &=& \fr{d_A}{2\pi^2}\bigg(\!\fr{g}{4\pi}\!\!\!\bigg)^{\!\!4}\sum_{fg}\mu_f^2\mu_g^2
\bigg\{\fr{7}{6} + \fr{1}{3}\fr{\mu_f^2+6\mu_g^2}{\mu_g^2}\ln\,\fr{\mu_f}{\mu_g} -
\fr{4}{3}\fr{\mu_f^2+\mu_g^2}{\mu_f\mu_g}\ln\, \fr{\mu_f+\mu_g}{\mu_g}
-\fr{(\mu_f-\mu_g)^4}{6\mu_f^2\mu_g^2}\ln\, \fr{\mid\!\mu_f^2-\mu_g^2\!\mid}{\mu_g^2}\bigg\}\nn
&=&\fr{d_A}{2\pi^2}\bigg(\!\fr{g}{4\pi}\!\!\!\bigg)^{\!\!4}\Bigg(\fr{7}{6}(\boldmu^2)^2-\fr{8}{3}\ln\,2\sum_f\mu_f^4
-\fr{1}{3}\sum_{f>g}\bigg\{(\mu_f-\mu_g)^2\ln\,\fr{|\mu_f^2-\mu_g^2|}{\mu_f\mu_g} \nn
&+& 4\mu_f\mu_g(\mu_f^2+\mu_g^2)\,\ln\,\fr{(\mu_f+\mu_g)^2}{\mu_f\mu_g}-
(\mu_f^4-\mu_g^4)\,\ln\,\fr{\mu_f}{\mu_g}\bigg\}\Bigg). \label{ppl2ares}
\ea
Adding Eqs. (\ref{ppl1res}), (\ref{p3b1}) and (\ref{ppl2ares}) together we finally obtain Eq. (\ref{pt03}) as the result for the whole plasmon
sum.



\end{document}